\documentclass[useAMS,usenatbib,usegraphicx]{mn2e}
\usepackage{amsmath}
\pdfoutput=1

\title[$\gamma$-ray observations of extraterrestrial neutrino track event positions]{$\gamma$-ray observations of extraterrestrial neutrino track events}

\author[A.M. Brown et al.]{Anthony M. Brown$^{1,2}$\thanks{E-mail: anthony.brown@durham.ac.uk}, Jenni Adams$^{2}$ and Paula M. Chadwick$^{1}$ \\
$^{1}$Department of Physics, University of Durham, South Road, Durham, DH1 3LE, UK \\
$^{2}$Department of Physics and Astronomy, University of Canterbury, Christchurch, NZ }
\begin{document}

\date{Accepted 2015 April 21. Received 2015 April 17. In original form 2015 January 25}

\pagerange{\pageref{firstpage}--\pageref{lastpage}} \pubyear{2015}

\maketitle

\label{firstpage}

\begin{abstract}
In this paper we report the results of a $\gamma$-ray study of IceCube's extraterrestrial neutrino candidates detected as track-like events. Using 70 months of \textit{Fermi}-LAT observations, a likelihood analysis of all $1-300$ GeV photons within 5\ensuremath{^{\circ}} of the track-like neutrino candidates' origin was undertaken, to search for spatially coincident $\gamma$-ray emission. One of IceCube's HESE track events was found to be spatially coincident with a $\gamma$-ray bright active galactic nucleus (AGN), PKS 0723-008. We find however, that the chance probability for \textit{Fermi}-LAT detected AGN to be spatially coincident with a single HESE track-like event is high ($\sim37$\%). We therefore find no evidence of $\gamma$-ray emission associated with the detection of IceCube's HESE track-like neutrino candidates. Upper limits were calculated in the energy range of $1-300$ GeV, assuming a point source origin for the neutrino events considered. The implications for the non-detection of $\gamma$-ray emission from the source of the HESE track-like events are briefly discussed. 

The large time period analysed in our study did however, reveal two new $\gamma$-ray point sources. With a flux of ($1.28 \pm 0.08$) $\times 10^{-9}$ photons cm$^{-2}$ s$^{-1}$, in the $1-300$ GeV energy range, and an associated TS value of 220.6, one of these new point sources is positionally coincident with the AGN PKS 1346-112. The other point source has a $1-300$ GeV flux of ($7.95 \pm 1.23$) $\times 10^{-10}$ photons cm$^{-2}$ s$^{-1}$ and an associated TS value of 92.4. This new point source is spatially coincident with the radio source NVSS J$072534+021645$ suggesting that it too is an AGN. 
\end{abstract}


\section{Introduction}
In 2013 the IceCube Neutrino Observatory published observational evidence for the existence of extraterrestrial high-energy neutrinos (\citet{icecube1}). This evidence was obtained through a `HESE'\footnote{IceCube's HESE analysis is its \textbf{H}igh \textbf{E}nergy \textbf{S}tarting \textbf{E}vent analysis, aimed at selecting high energy events that have an interaction vertex within the IceCube detector.} analysis of the first two years of data taken with the 79 and 86 string IceCube detector. The addition of a third year of data resulted in the evidence evolving to a 5 $\sigma$ discovery of these extraterrestrial neutrinos (\citet{icecube2}). IceCube's 3-year HESE analysis uncovered a total of 37 extraterrestrial neutrino candidates with deposited energy in the range of 30 TeV to 2 PeV. The null hypothesis that all of these 37 events are from an atmospheric background was rejected at the $5.7\sigma$ level of confidence. Employing a series of likelihood-based tests, the IceCube collaboration found no significant evidence against an isotropic distribution for the HESE neutrino candidates. With the discovery of extraterrestrial neutrinos, IceCube has open a new window through which to view the high-energy Universe. Due to their importance, these 37 candidate neutrino events have been interpreted within a range of galactic and extra-galactic scenarios (e.g. \citet{razzaque}, \citet{padovani}, \citet{dermer}, \citet{dermer2}, \citet{krauss}). 

Neutrino emission from high-energy astrophysical sources capable of accelerating protons to relativistic energies has long been predicted. The neutrinos arise from the decay of charged mesons created in the interactions of the relativistic protons with either ambient gas (p-p) or ambient radiation (p-$\gamma$). If the 20 TeV $-$ 2 PeV extraterrestrial neutrinos detected by IceCube were indeed created this way, then their existence implies the existence of astrophysical objects capable of accelerating protons to energies of at least O(10) to O(100) PeV. The decay of neutral mesons created in the same interactions of the relativistic protons produce $\gamma$-rays. While the p-p interactions produce a smooth power law distribution for the resultant neutrinos and $\gamma$-ray fluxes, the p-$\gamma$ interactions usually produce a peaked distribution for the resultant neutrinos and $\gamma$-ray photons, with the peak and spread of these distributions strongly dependent upon the nature of the `target' photon population (M\"{u}cke et al. (2000a,b)). Furthermore, the ratio of the typical neutrino and $\gamma$-ray energy, $E_{\gamma}/E_{\nu}$, also depends upon the characteristics of the target photons, with a ratio of $\sim1.2$ expected for a synchrotron radiation target and $\sim1.1$ for a thermal radiation target distribution. Regardless of whether the interaction is synchrotron p-$\gamma$ or thermal p-$\gamma$, the $\sim10-100$s TeV $\gamma$-rays associated with the $\sim30-250$ TeV neutrino candidates considered in this study are believed to quickly cascade down to $E_{\gamma}<100$ GeV energies due to the photon opacity of the emission region (\citet{mannheim}). Assuming that the opacity of the neutrino$/ \gamma$-ray emission region is large enough to facilitate the transformation of the initial TeV photons down to GeV energies while not being so large as to absorb the $\gamma$-ray flux entirely, a search for the neutrino sources can be made with the \textit{Fermi}-LAT detector.
 
The LAT detector onboard the \textit{Fermi} satellite, described in detail by \citet{atwood}, is a pair-conversion telescope, sensitive to a photon energy range from below 20 MeV to above 300 GeV. With a large field of view, $ \simeq 2.4 $ sr, improved angular resolution, 68\% containment angle of $\sim0.8$\ensuremath{^{\circ}} at 1 GeV, and large effective area, $\sim 8000$ cm$^{2}$ on axis for 10 GeV photons, \textit{Fermi}-LAT provides an order of magnitude improvement in performance compared to its \textit{EGRET} predecessor. Since 2008 August 4, the vast majority of data taken by \textit{Fermi} has been performed in \textit{all-sky-survey} mode, whereby the \textit{Fermi}-LAT detector points away from the Earth and rocks north and south of its orbital plane, on subsequent orbits. This rocking motion, coupled with \textit{Fermi}-LAT's large effective area, allows \textit{Fermi} to scan the entire $\gamma$-ray sky every two orbits, or approximately every three hours. This observational characteristic affords us, for the first time, continuous monitoring of the high-energy $\gamma$-ray sky and thus has important implications for the study of transient events. In addition to this, coupling \textit{Fermi}-LAT's continual scanning of the sky with a long mission lifetime allows us to construct a deep exposure of the $\gamma$-ray sky. 

IceCube classifies events by the pattern of the Cherenkov light seen in the detector array; {\em track} events are those where the presence of a high-energy muon has produced a distinctive visible track of light while {\em shower} events are those where the Cherenkov light, from the particle shower initiated when the neutrino interacts, is observed only in aggregate. The angular resolution for events containing visible muon tracks is much better ($\leq 1$\ensuremath{^{\circ}}) than for those that do not ($\sim 15$\ensuremath{^{\circ}} (\citet{icecube2}). For equal neutrino fluxes of all flavours, events with tracks make up only 20\% of interactions (\citet{icecube2}). Eight of the 37 HESE neutrino candidate events were classified as track events, 28 as shower events and one event was formed by two coincident cosmic-ray muon events. Although there are fewer track events, the angular resolution of the track events is comparable with that of the \textit{Fermi}-LAT detector for $E_{\gamma}>1$ GeV photons which facilitates associating the track-like events of IceCube’s HESE analysis with known \textit{Fermi}-LAT detected $E_{\gamma}>1$ GeV $\gamma$-ray sources. 

An important product of \textit{Fermi}-LAT's continual scanning of the sky is the $\gamma$-ray catalogues that it periodically produces such as the one and two year point source catalogues (1FGL (\citet{gamma}) and 2FGL (\citet{nolan})), or the second Active Galaxies catalogues (2LAC (\citet{acker3})). The \textit{Fermi}-LAT collaboration recently released their first high-energy catalog (1FHL), which lists all $E_{\gamma}>10$ GeV sources from the first three years of \textit{Fermi}-LAT observations (\citet{1fhl}). 

\citet{padovani} looked for astrophysical counterparts to the $E_{\nu}>60$ TeV HESE neutrino candidates in three $\gamma$-ray catalogues including the 1FHL catalogue. Padovani \& Resconi placed their highest priority on the TeVCat catalogue which reaches the TeV regime. Unsurprisingly, given that the neutrino sample of their study was dominated by shower events (83\%) with an average median angular error of $\sim11.4$\ensuremath{^{\circ}}, they found a large amount of spatial correlation between the catalogues and their neutrino candidates. To therefore determine the plausibility of the association between the TeV emitting coincident objects, they had identified, and the neutrino candidates, Padovani \& Resconi compared the $\gamma$-ray spectral energy density of the object with the neutrino flux per event at the specific energy of the relevant neutrino and assessed whether a simple extrapolation plausibly connected the $\gamma$-rays with the neutrino. Around 10~\% of the TeVCat counterparts passed this energy diagnostic. For the counterparts which were listed only in the 1FHL catalogue, Padovani \& Resconi decided there was too big a gap between the source's highest detected $\gamma$-ray energy and the neutrino energy to make a sensible link. Padovani \& Resconi found no spatial correlation between the 1FHL catalogue and the median angular error radius of the three track-like events in their neutrino sample, although they did note that the 2FGL source PKS 0723-008 is correlated with neutrino candidate \#5 and that the 1FHL source 4C$+41.11$ and the 2FGL source PKS 0823-321 were just outside the median angular errors of candidates \#13 and \#3 respectively.

In what could be considered the inverse cross-correlation, the IceCube collaboration have recently performed a likelihood search for cumulative neutrino emission from a large sample of of 2LAC sources (\citet{glusen}). The neutrino sample used was a muon-track sample from the years 2009-2011 containing over 300,000 events with a median angular resolution of 1\ensuremath{^{\circ}} at 1 TeV and 0.5\ensuremath{^{\circ}} at 100 TeV. In the southern hemisphere the sample is dominated by muons from air showers and in the upgoing region the majority of events come from atmospheric muon neutrinos. This was accounted for in the likelihood analysis through the background probability distribution function. The study found no significant correlation and concluded that 2LAC blazars are not responsible for the majority of the diffuse astrophysical neutrinos reported in \citet{icecube2}.

Thus, there is currently, no clear association between the IceCube extraterrestrial neutrino candidates and $\gamma$-ray sources and as Padovani \& Resconi remarked, this strongly motivates more TeV photon observations. If it is the case, as mentioned above, that the $\sim10-100$s TeV $\gamma$-rays  quickly cascade down to $E_{\gamma}<100$ GeV energies then the $\gamma$-ray counterparts of the neutrino sources may be revealed through a close analysis of \textit{Fermi}-LAT observations.  Here, rather than cross-correlating against precompiled catalogues, we use the deep exposure of the first 70 months of \textit{Fermi}-LAT operation to search for GeV counterparts to the track events of IceCube's 3-year HESE analysis. Such an analysis, while sensitive to catalogued sources, also allows us to search for faint $\gamma$-ray sources not present in the existing \textit{Fermi} catalogues and, in the absence of evidence for a source, determine upper limits for the associated $\gamma$-ray flux. Also the 70 month \textit{Fermi}-LAT data set contains the 3 year time period covered by the IceCube HESE search which allows our analysis to be sensitive to the possibility that the neutrinos were produced during the flare of a normally faint source.  All neutrino candidates with a track-like topology are investigated, with the exception of event \#28, which displays evidence of a cosmic ray induced air shower in the IceTop detector accompanying the neutrino candidate, thus suggesting it is part of the atmospheric background. Each of the remaining neutrino candidates are considered independently. In Section 2, we describe the \textit{Fermi}-LAT observations and data analysis routines employed during our study. Section 3 touches on the GeV properties of the new $\gamma$-ray point sources discovered during our study, with Section 4 briefly discussing possible interpretations of the non-detection of GeV emission spatially co-incident with the origin of the neutrino candidates considered. 

\section{\textit{Fermi}-LAT observations and data analysis}
The data used in this study comprise all \textit{Fermi}-LAT event and spacecraft data taken during the first 70 months of \textit{Fermi}-LAT operation, from 2008 August 4 to 2014 June 4, which equates to a Mission Elapsed Time (MET) interval of 239557417 to 423589417. All \textsc{source} $\gamma$-ray events\footnote{\textsc{source} events have an event class of 2 in the \textsc{pass}7\_\textsc{rep} data, and have a high probability of being a $\gamma$-ray (see \citet{acker1} for details of event classification).}, in the $1 < E_{\gamma} < 300$ GeV energy range, within a 5\ensuremath{^{\circ}} radius of interest (RoI) centered on the right ascension and declination of the origin of HESE track-like events were considered\footnote{The right ascension and declination for IceCube's HESE track events were taken from Table 1 of \citet{icecube2}.}. In accordance with the \textsc{pass}7\_\textsc{rep} criteria, a zenith cut of 100\ensuremath{^{\circ}} was applied to the data to remove any cosmic ray induced $\gamma$-rays from the Earth's atmosphere. The good time intervals were generated by applying a filter expression of ``\textsc{(data\_qual$==$1) \&\& (lat\_config$==$1) \&\& abs(rock\_angle)$<$ 52}'' to the data\footnote{See \url{http://fermi.gsfc.nasa.gov/ssc/data/analysis/} \url{documentation/Cicerone/Cicerone_Data_Exploration/} \url{Data_preparation.html} for details on LAT data selection.}. A summary of the analysis criteria is given in Table \ref{analysis}. 

\begin{table}
   \caption{Summary of the criteria utilised in this analysis.}
     \begin{tabular}{ll} \hline \hline 
      Science Tools version     & \textsc{v9r33p0}  \\ 
      IRF                       & \textsc{p7rep\_source\_v15}    \\ 
      Event class               & \textsc{source}, Reprocessed Pass 7     \\
      Photon Energies           & $1-300$ GeV    \\ 
      RoI                       & 5\ensuremath{^{\circ}}      \\
      Zenith angle cut          & $<100$\ensuremath{^{\circ}}    \\  
      Rocking angle cut         & $<52$\ensuremath{^{\circ}}    \\
      LAT config/Data quality   & $==1$   \\
      Galactic diffuse model    & gll\_iem\_v05\_rev1.fit \\
      Isotropic diffuse model   & iso\_source\_v05.txt \\ \hline \hline
    \end{tabular}
  \label{analysis}
\end{table}

Throughout this analysis, version \textsc{v9r33p0} of the \textit{Fermi Science Tools} was used in conjunction with the \textsc{p7rep\_source\_v15} instrument response functions (IRFs). During the likelihood analysis, a model file consisting of both point and diffuse sources of $\gamma$-rays was employed. For the diffuse emission, the most recent Galactic, gll\_iem\_v05\_rev1.fit, and extragalactic, iso\_source\_v05.txt, diffuse models were used. In addition to these diffuse models, all point sources within 10\ensuremath{^{\circ}} of the candidate neutrino position ($\alpha_{J2000}$, $\delta_{J2000}$) were considered in the likelihood model. The position and spectral shape for these point sources were taken from the 2FGL. The normalisations and spectral indices of all point sources in the model file were left free to vary, with the exception of those within the 5\ensuremath{^{\circ}} to 10\ensuremath{^{\circ}} annulus from the candidate neutrino event ($\alpha_{J2000}$, $\delta_{J2000}$) position which were fixed to their 2FGL catalogue values. The normalisation factor of the extragalactic diffuse emission was left free to vary, while the Galactic diffuse template was multiplied by a power law in energy, the normalisation of which was also left free to vary. The model map immediately allows us to determine if there are any catalogued sources spatially coincident with the neutrino candidates and, as noted by \citet{padovani}, there is one such coincidence with the  2FGL AGN  located within median angular error of  neutrino candidate \#5. The significance of this spatial correlation, along with the observed $\gamma$-ray properties over the 70-months, are investigated in detail in \textsection 2.1.

An initial \textsc{binned} likelihood analysis was conducted to confirm the accuracy of the `diffuse $+$ 2FGL point source' model. 
The best-fit model from this initial likelihood fit was used, in conjunction with the \textit{Fermi} tool \textsc{gtmodel}, to construct a `model map' of a 7\ensuremath{^{\circ}}$\times$7\ensuremath{^{\circ}} region centered on the ($\alpha_{J2000}$, $\delta_{J2000}$) of each candidate neutrino track event. This model map was then compared with the 70-month sky map of the $1-300$ GeV $\gamma$-rays to create a residuals count map\footnote{The residuals count map was constructed via (sky map$-$model map)$/$model map (e.g. see \citet{meVHE2})}. Any positive excess of events in the residuals map indicates the presence of a new source of $\gamma$-rays\footnote{The use of a residuals map to look for new $\gamma$-ray sources is independent and complementary to the alternative approach of using the  \textsc{gttsmap} \textit{Fermi} tool (see eg. \citet{mepicA}, \citet{oscar} \& \citet{meVHE1}).}. If new sources are present in the data, and not properly accounted for within the model file utilised during the likelihood analysis, they can artifically increase the significance of the $\gamma$-ray flux from the neutrino candidates. The 7\ensuremath{^{\circ}}$\times$7\ensuremath{^{\circ}} sky map, model map and residuals maps for the 7 separate HESE track events can be seen in Figures 1 to 7. All maps have been smoothed with a 1\ensuremath{^{\circ}} Gaussian. The colour scales for the sky and model maps are in units of $\gamma$-ray counts, while the residuals maps are in units of percentage. 

\begin{figure*}
 \begin{minipage}{150mm}
  \centering

\includegraphics[angle=90,width=.33\textwidth]{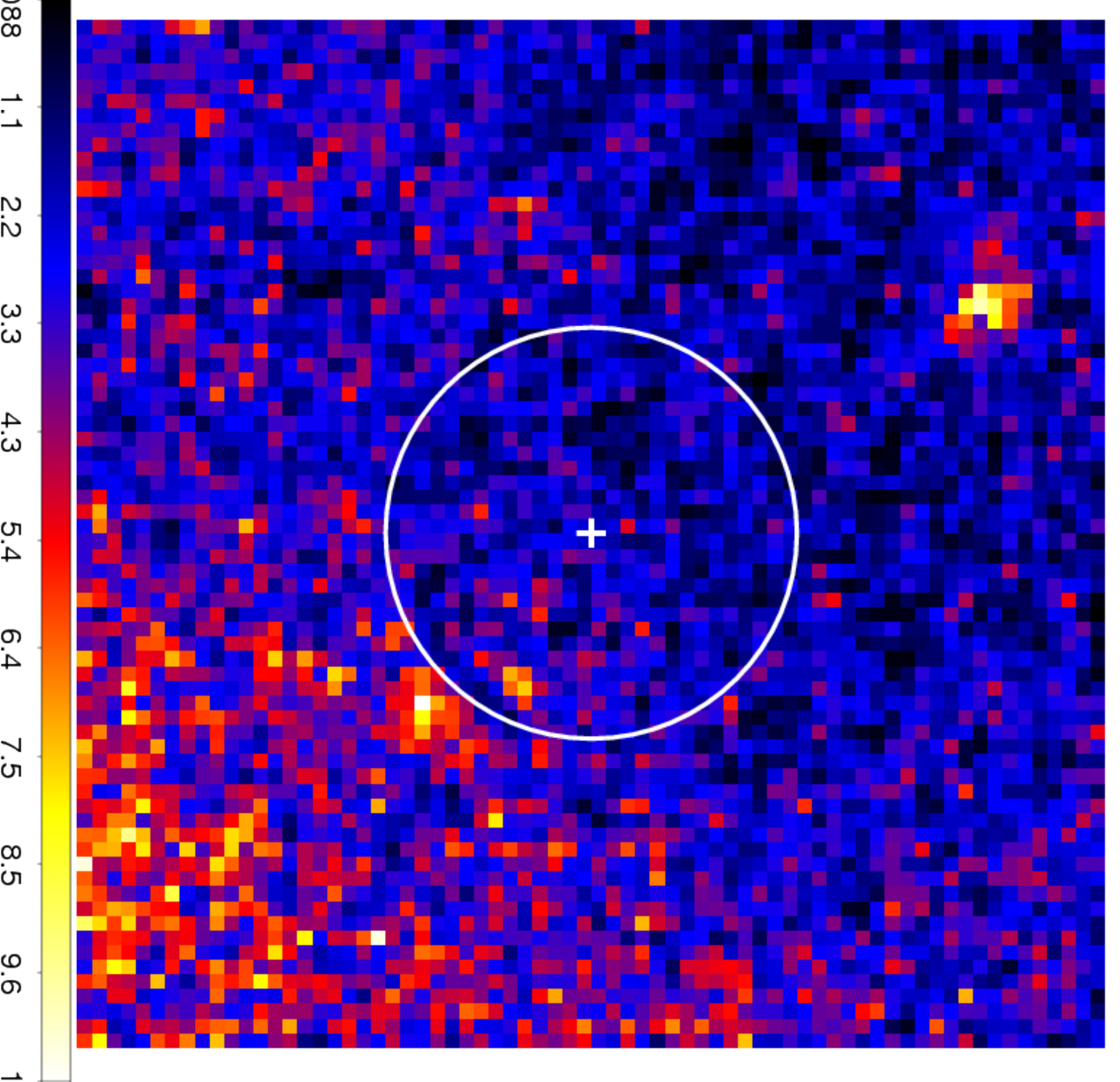}\hfill
\includegraphics[angle=90,width=.33\textwidth]{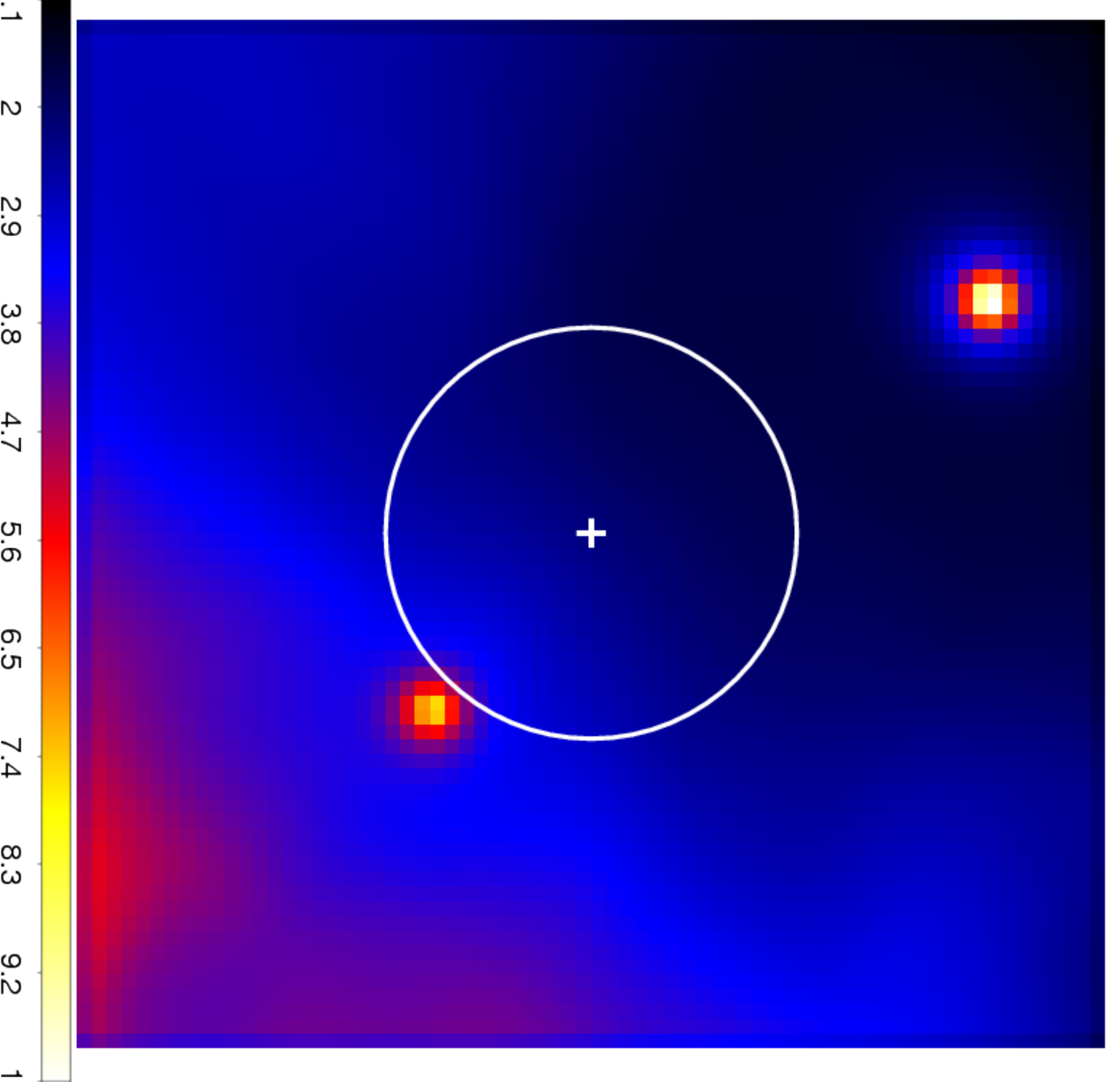}\hfill
\includegraphics[angle=90,width=.33\textwidth]{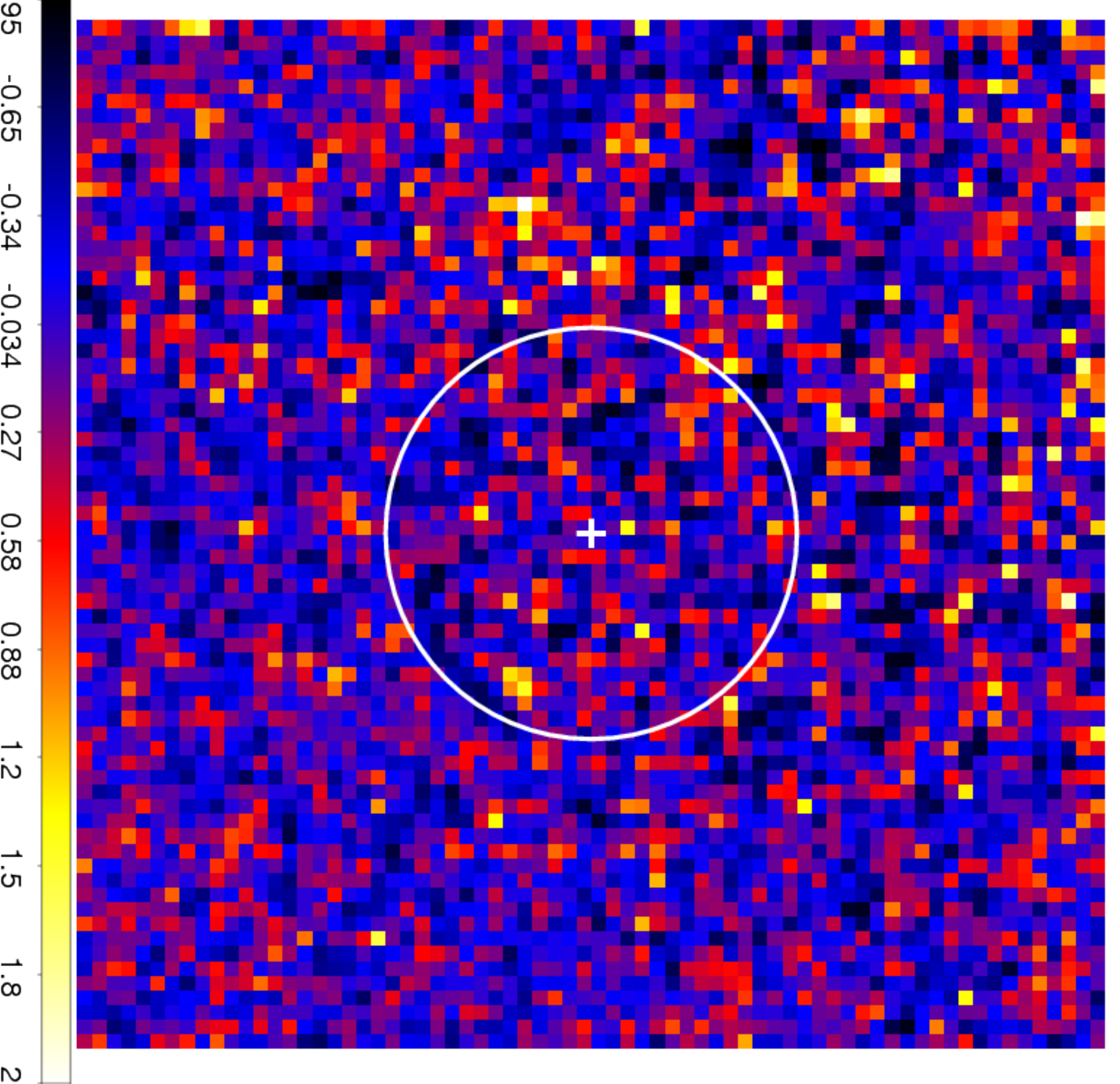}

\caption{Event 3: Left: 7\ensuremath{^{\circ}}$\times$7\ensuremath{^{\circ}} sky map of the $1-300$ GeV flux for the entire 5.8 year data set. Middle: 7\ensuremath{^{\circ}}$\times$7\ensuremath{^{\circ}} model map produced with the \textit{Fermi} tool \textsc{gtmodel}, utilising the best-fit model from the 5.8 year binned likelihood analysis.  Right: 7\ensuremath{^{\circ}}$\times$7\ensuremath{^{\circ}} residuals map of the $1-300$ GeV events. The residuals map is produced by (sky map$-$model map)$/$model map). All maps have been smoothed with a 1\ensuremath{^{\circ}} Gaussian. The cross at the centre of each map indicates the ($\alpha_{J2000}$, $\delta_{J2000}$) of the neutrino candidate, while the circle present in each map indicates the median error radius for the neutrino candidate as shown in Table 2. The colour scales for the sky and model maps are in units of $\gamma$-ray counts, while the residuals maps are in units of percentage.There is no evidence for new $\gamma$-ray sources beyond the diffuse $+$ 2FGL point sources model used.}
 \label{fig:figure3}
 \end{minipage}
\end{figure*}

\begin{figure*}
 \begin{minipage}{150mm}
  \centering

\includegraphics[angle=90,width=.33\textwidth]{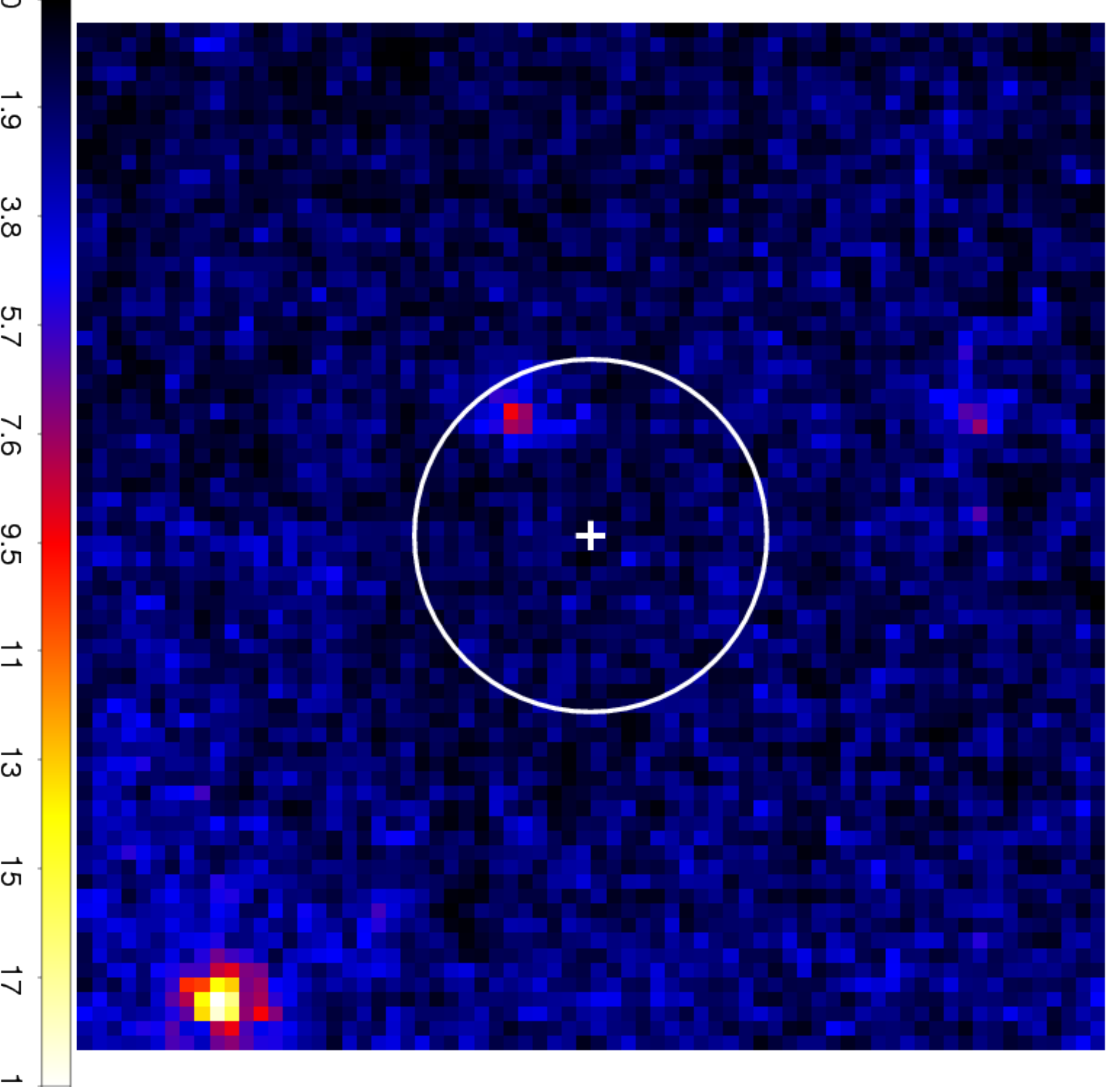}\hfill
\includegraphics[angle=90,width=.33\textwidth]{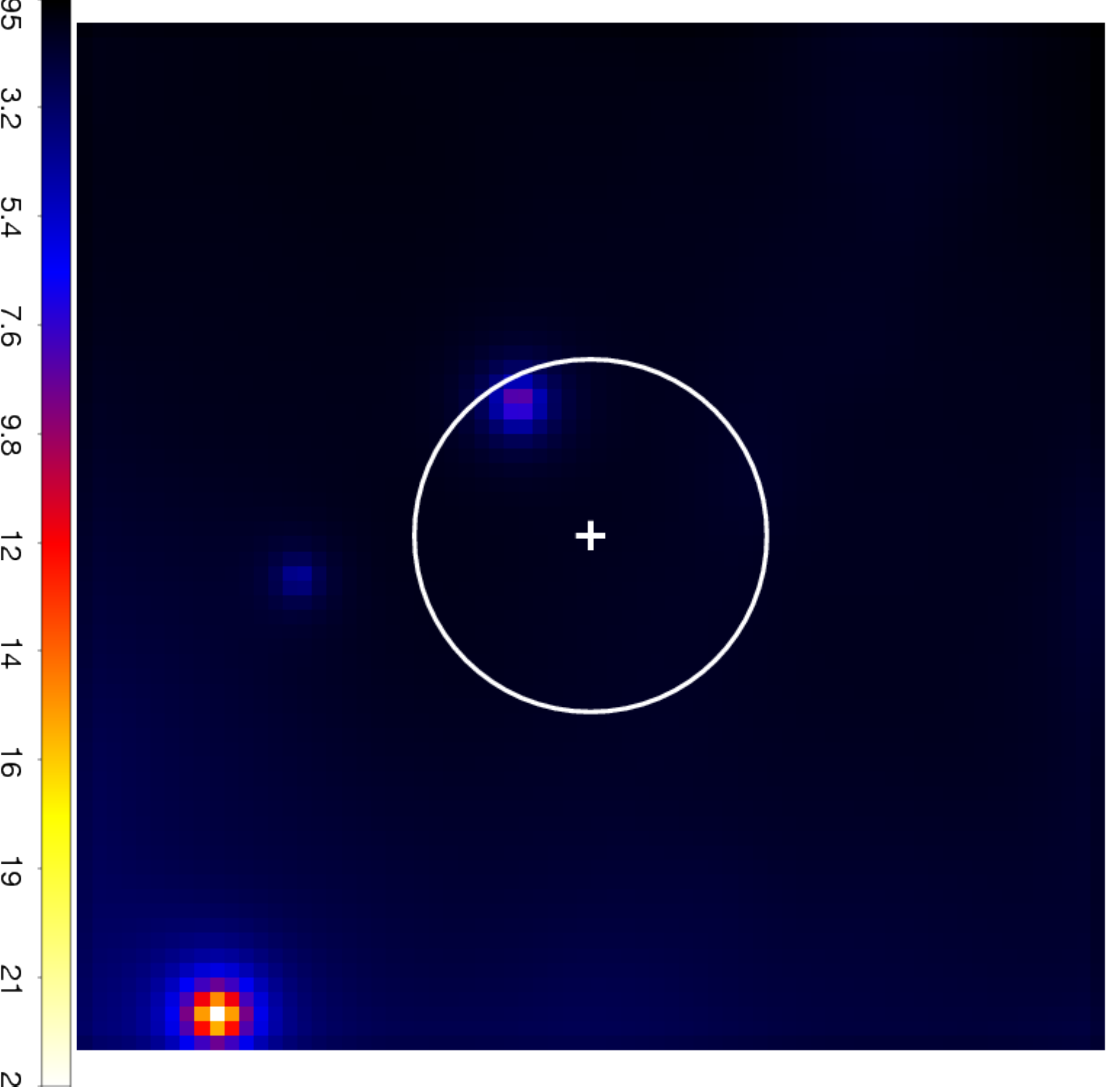}\hfill
\includegraphics[angle=90,width=.33\textwidth]{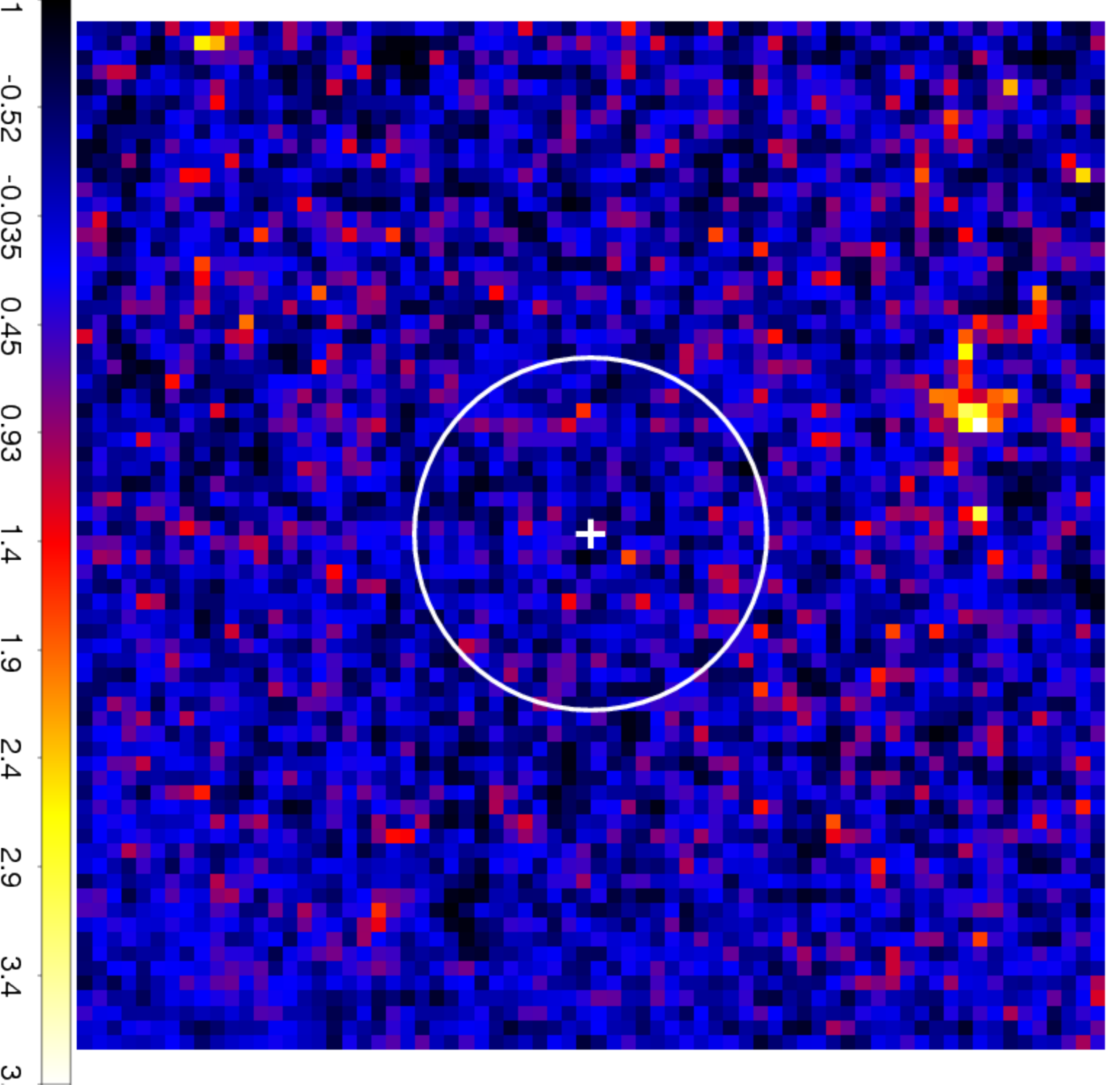}

\caption{7\ensuremath{^{\circ}}$\times$7\ensuremath{^{\circ}} sky, model and residuals maps for Event 5. There is an excess in the residuals map consistent with a point source at ($\alpha_{J2000}$, $\delta_{J2000}$) $=$ (111.419, 2.221).}
  \label{fig:figure3}
 \end{minipage}
\end{figure*}

\begin{figure*}
 \begin{minipage}{150mm}
  \centering

\includegraphics[angle=90,width=.33\textwidth]{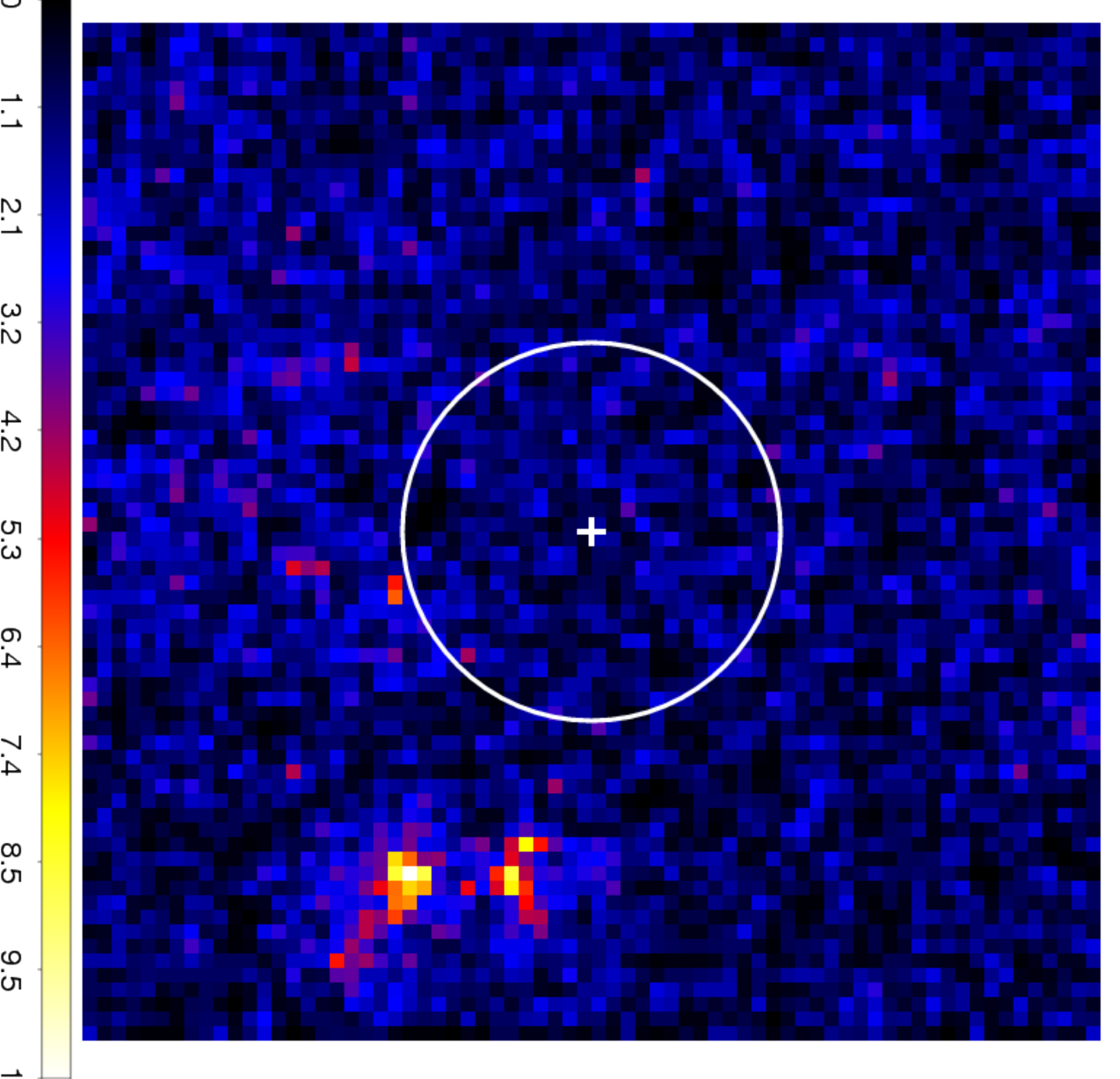}\hfill
\includegraphics[angle=90,width=.33\textwidth]{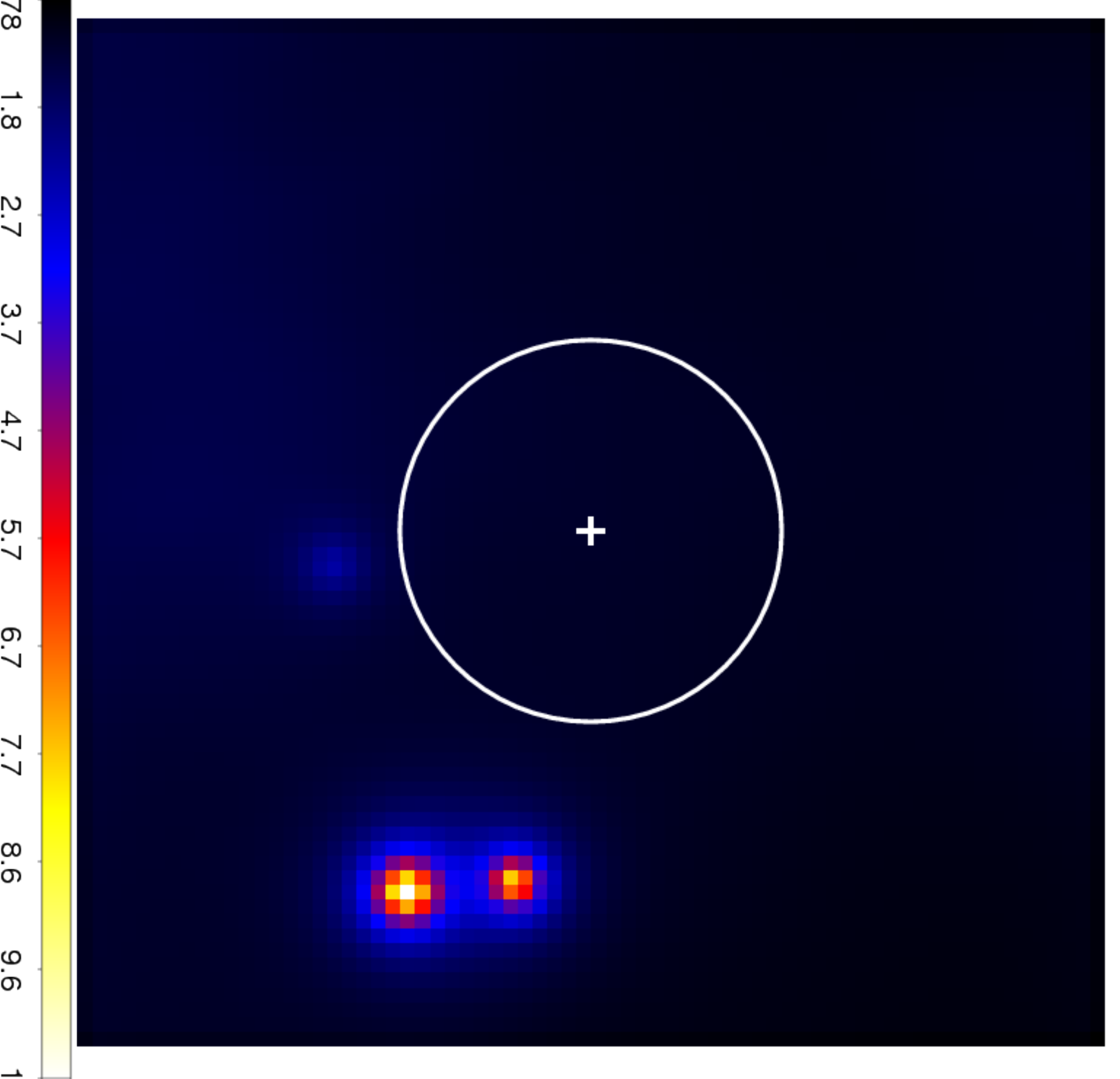}\hfill
\includegraphics[angle=90,width=.33\textwidth]{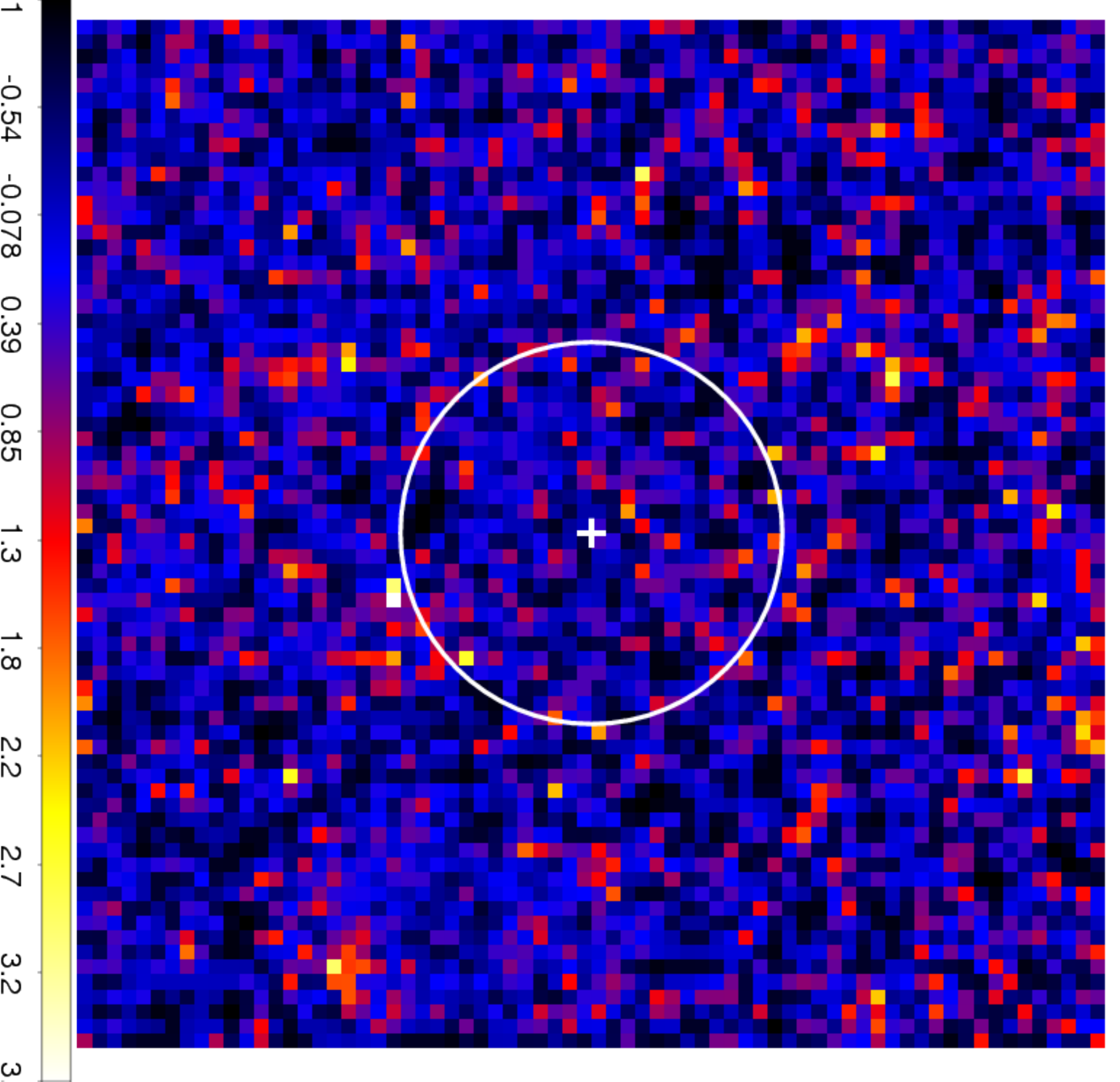}

\caption{7\ensuremath{^{\circ}}$\times$7\ensuremath{^{\circ}} sky, model and residuals maps for Event 8. There is no evidence for new $\gamma$-ray sources beyond the diffuse $+$ 2FGL point sources model used.}
 \label{fig:figure3}
 \end{minipage}
\end{figure*}

\begin{figure*}
 \begin{minipage}{150mm}

\centering
\includegraphics[angle=90,width=.33\textwidth]{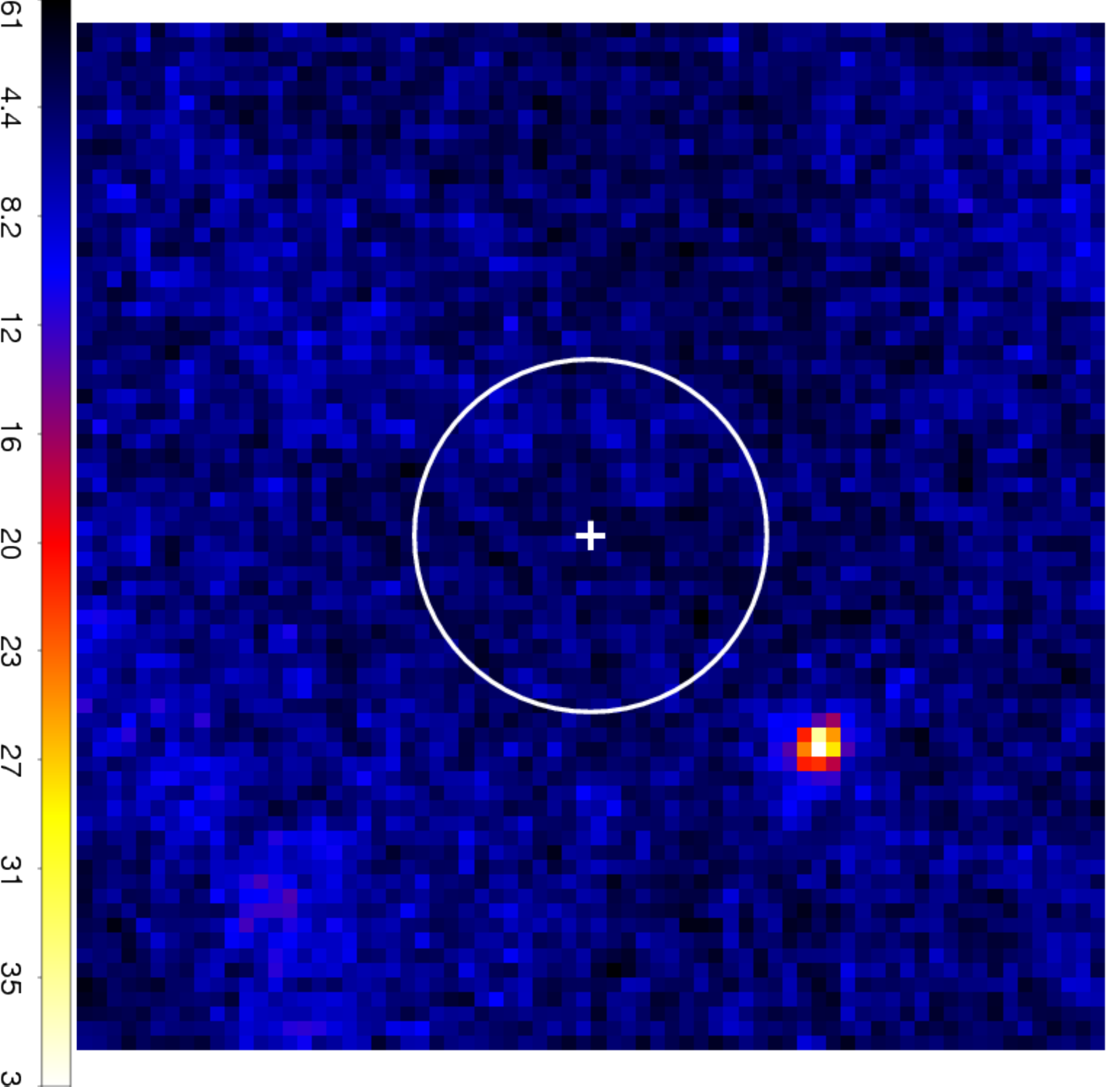}\hfill
\includegraphics[angle=90,width=.33\textwidth]{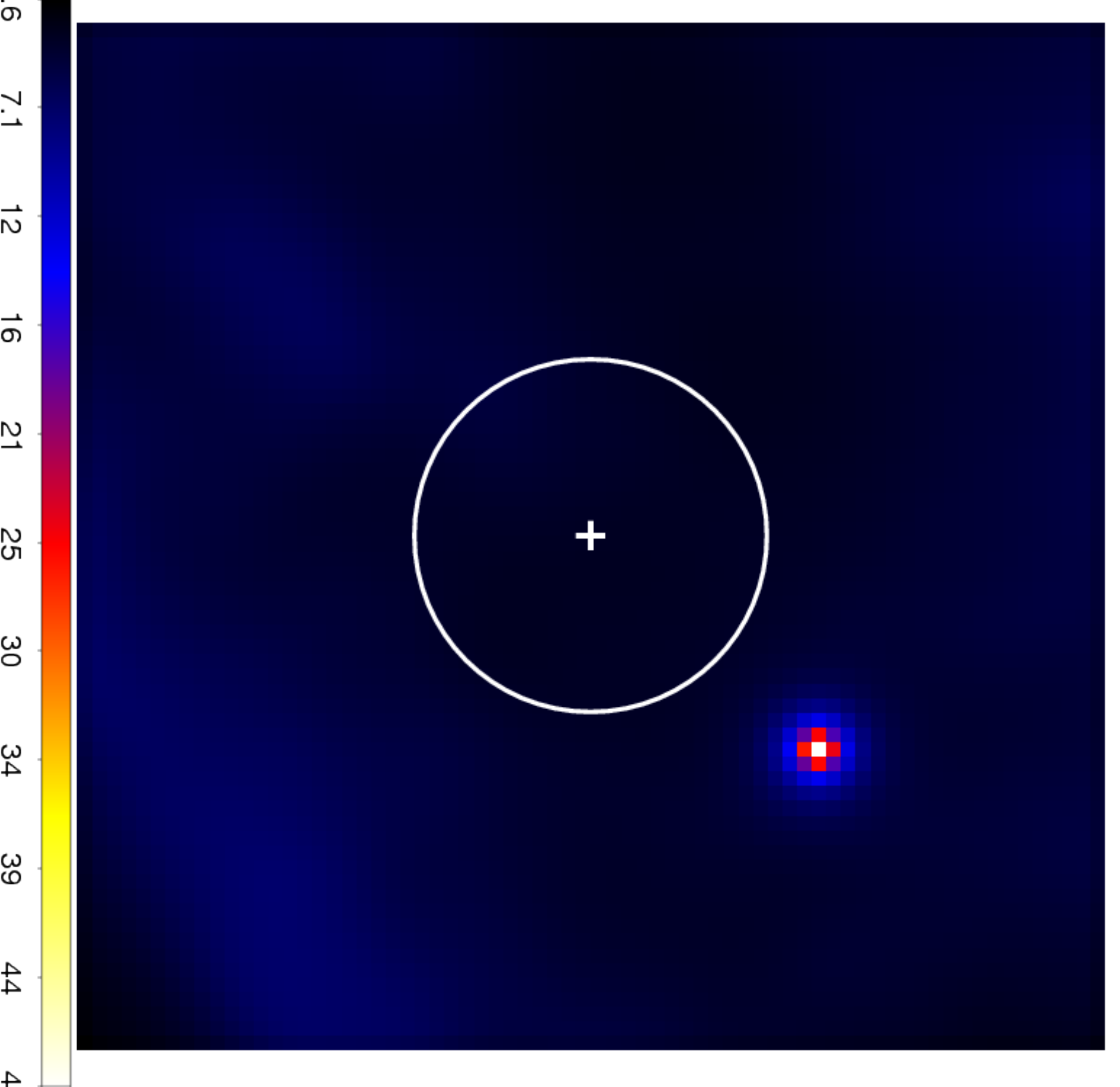}\hfill
\includegraphics[angle=90,width=.33\textwidth]{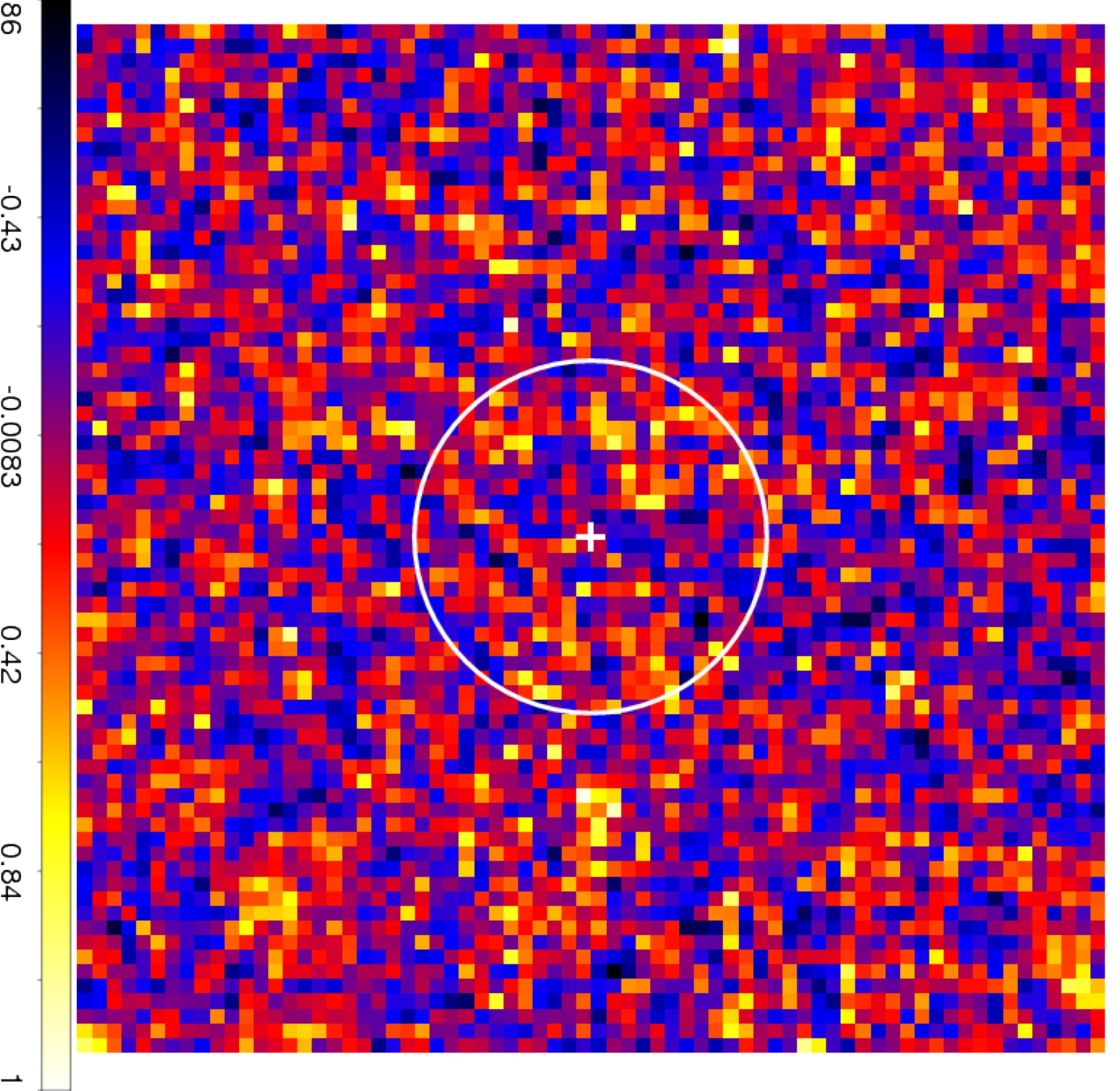}

\caption{7\ensuremath{^{\circ}}$\times$7\ensuremath{^{\circ}} sky, model and residuals maps for Event 13. There is no evidence for new $\gamma$-ray sources beyond the diffuse $+$ 2FGL point sources model used.}
\label{fig:figure3}

 \end{minipage}
\end{figure*}

\begin{figure*}
 \begin{minipage}{150mm}
  \centering

\includegraphics[angle=90,width=.33\textwidth]{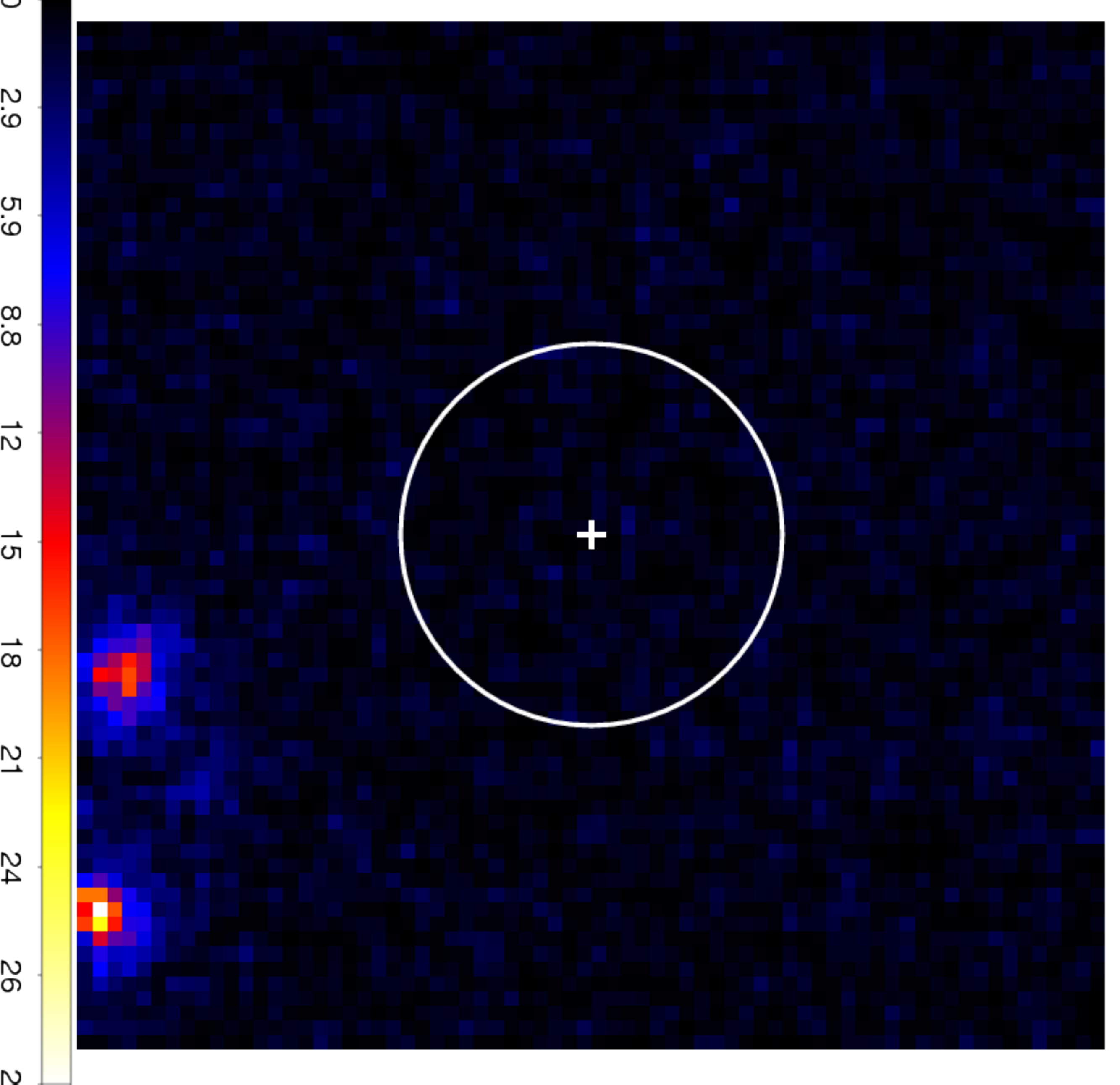}\hfill
\includegraphics[angle=90,width=.33\textwidth]{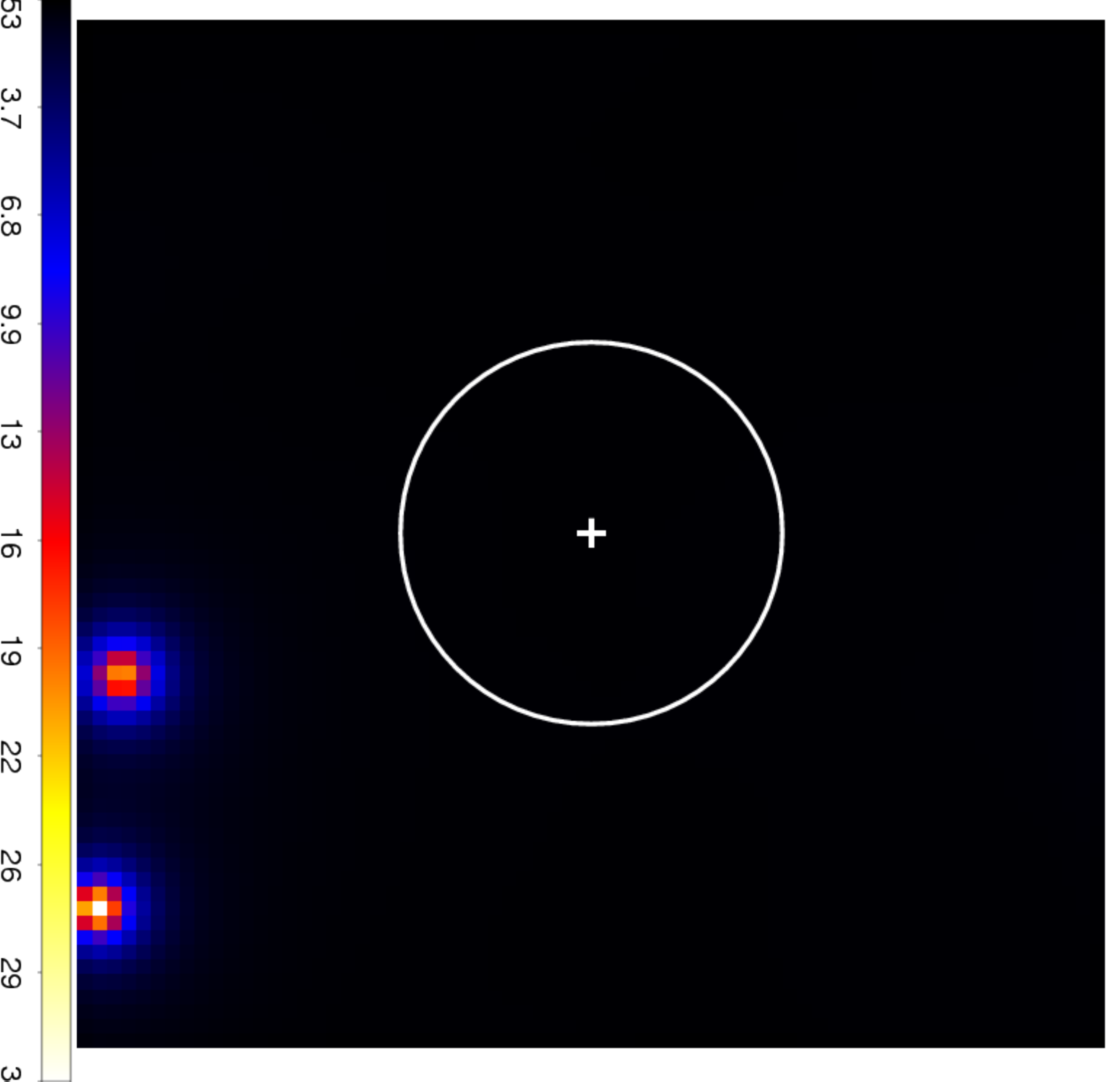}\hfill
\includegraphics[angle=90,width=.33\textwidth]{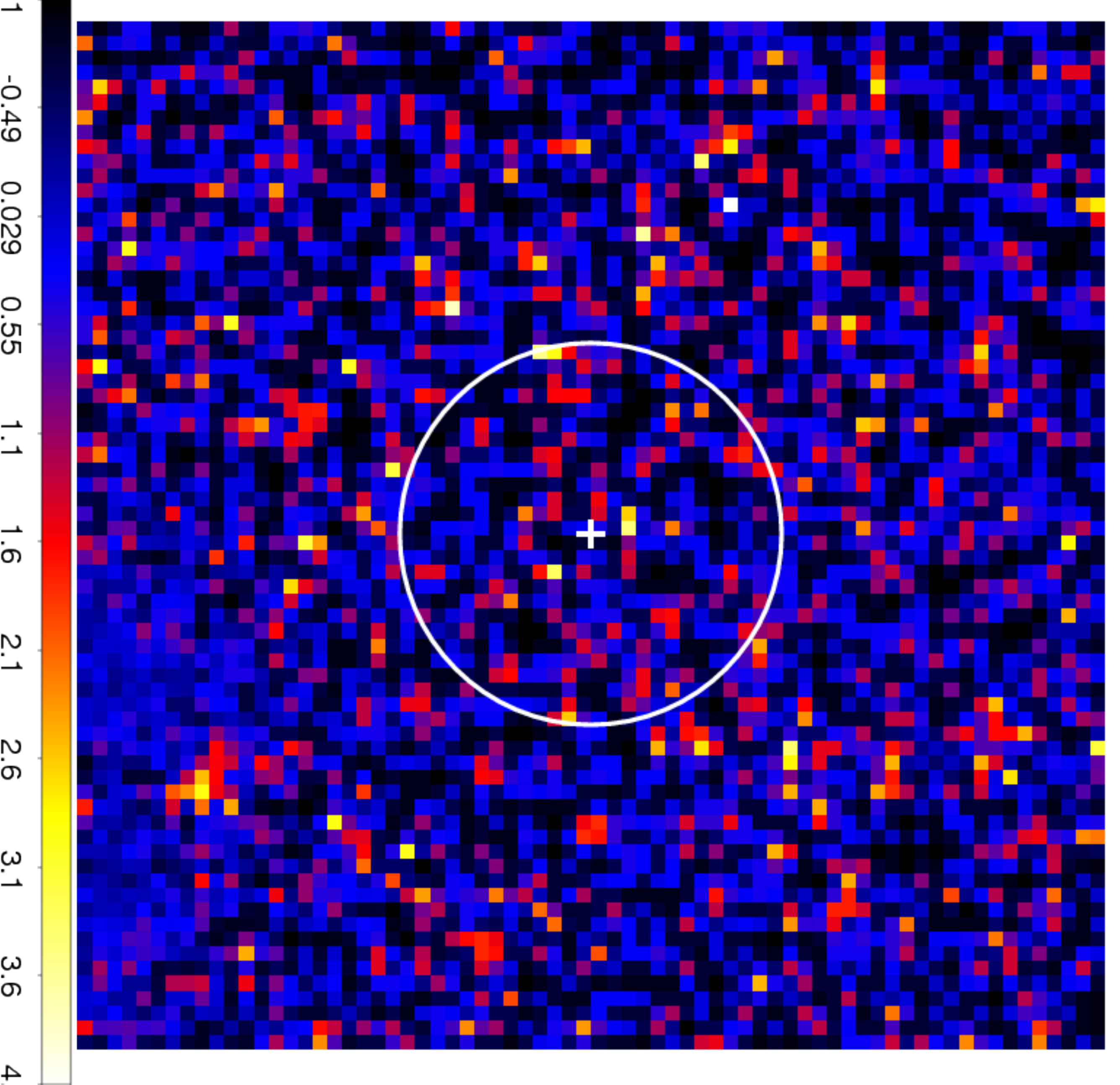}

\caption{7\ensuremath{^{\circ}}$\times$7\ensuremath{^{\circ}} sky, model and residuals maps for Event 18. There is no evidence for new $\gamma$-ray sources beyond the diffuse $+$ 2FGL point sources model used.}
 \label{fig:figure3}
 \end{minipage}
\end{figure*}

\begin{figure*}
 \begin{minipage}{150mm}
  \centering

\includegraphics[angle=90,width=.33\textwidth]{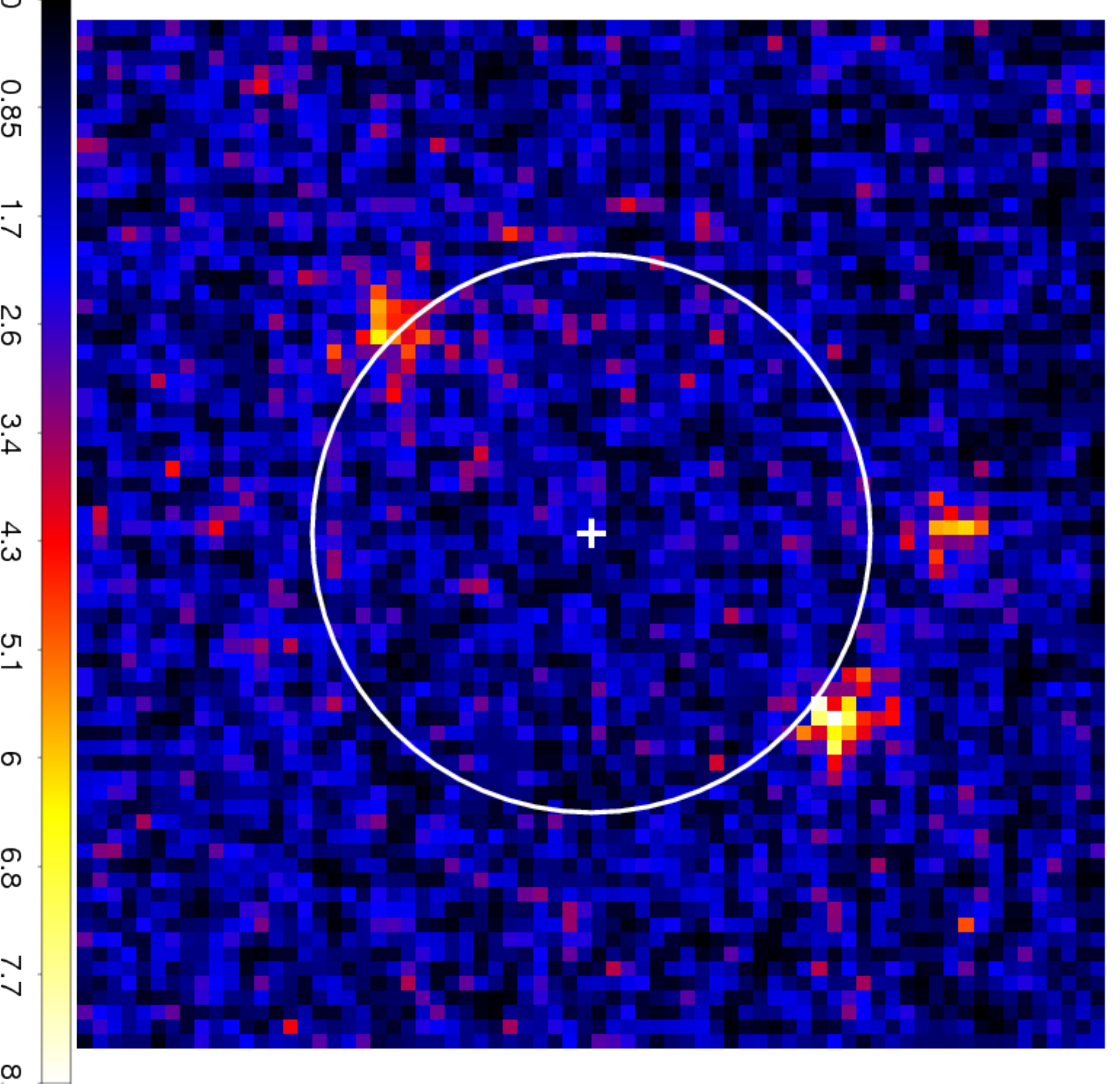}\hfill
\includegraphics[angle=90,width=.33\textwidth]{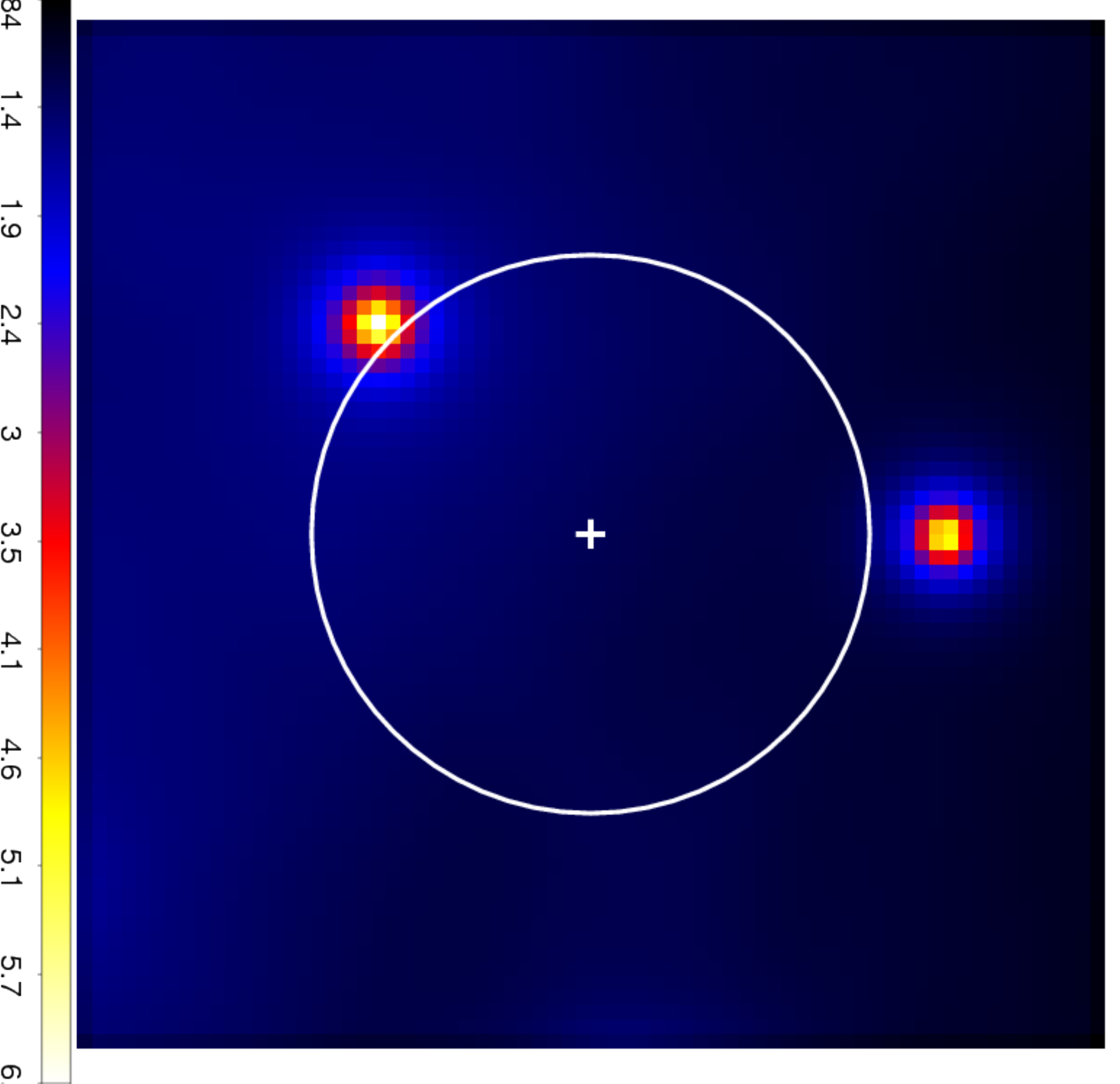}\hfill
\includegraphics[angle=90,width=.33\textwidth]{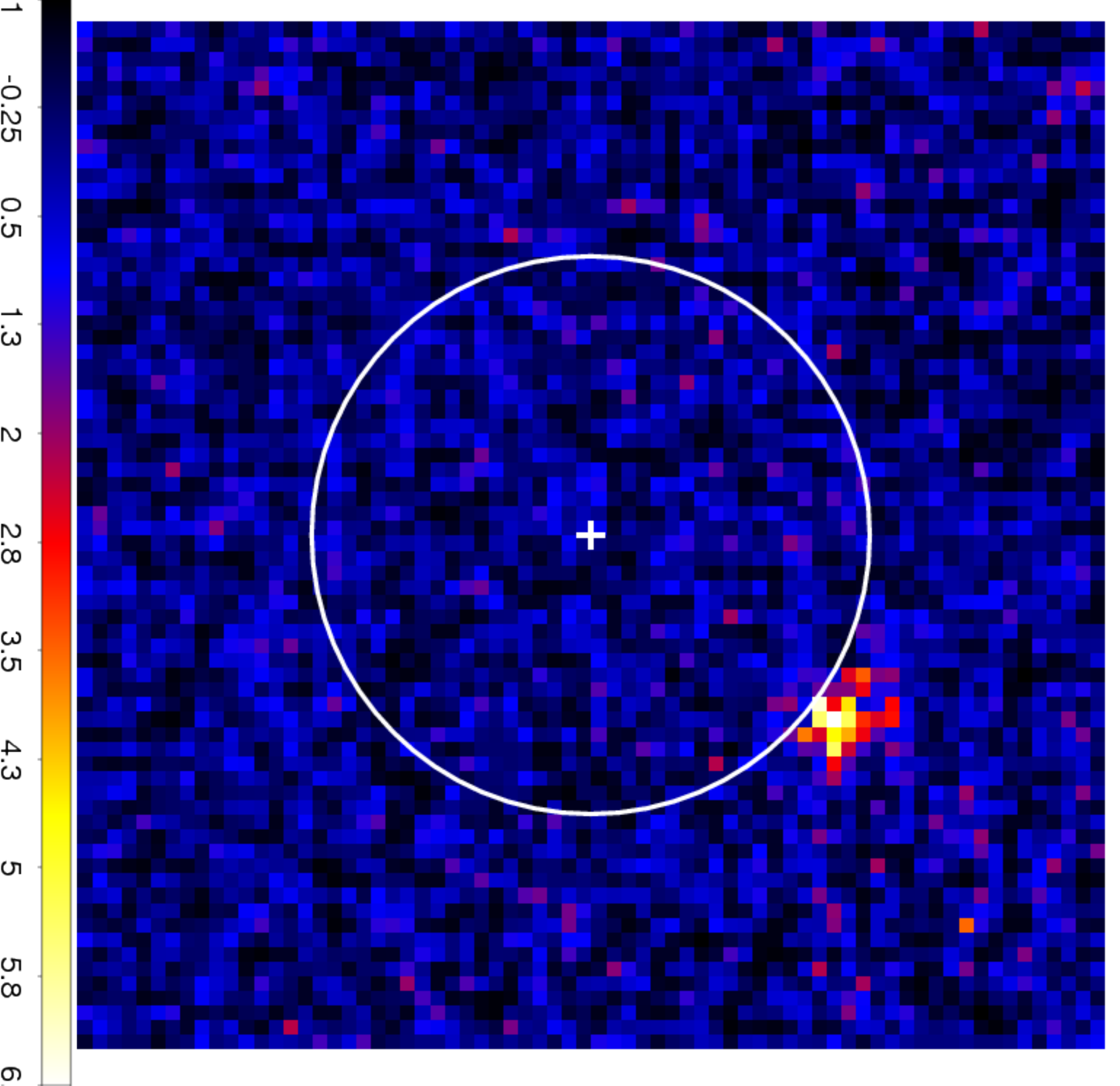}

\caption{7\ensuremath{^{\circ}}$\times$7\ensuremath{^{\circ}} sky, model and residuals maps for Event 23. There is a excess in the residuals map consistent with a point source at ($\alpha_{J2000}$, $\delta_{J2000}$) $=$ (207.409,-11.553).}
  \label{fig:figure3}
 \end{minipage}
\end{figure*}

\begin{figure*}
 \begin{minipage}{150mm}

\centering
\includegraphics[angle=90,width=.33\textwidth]{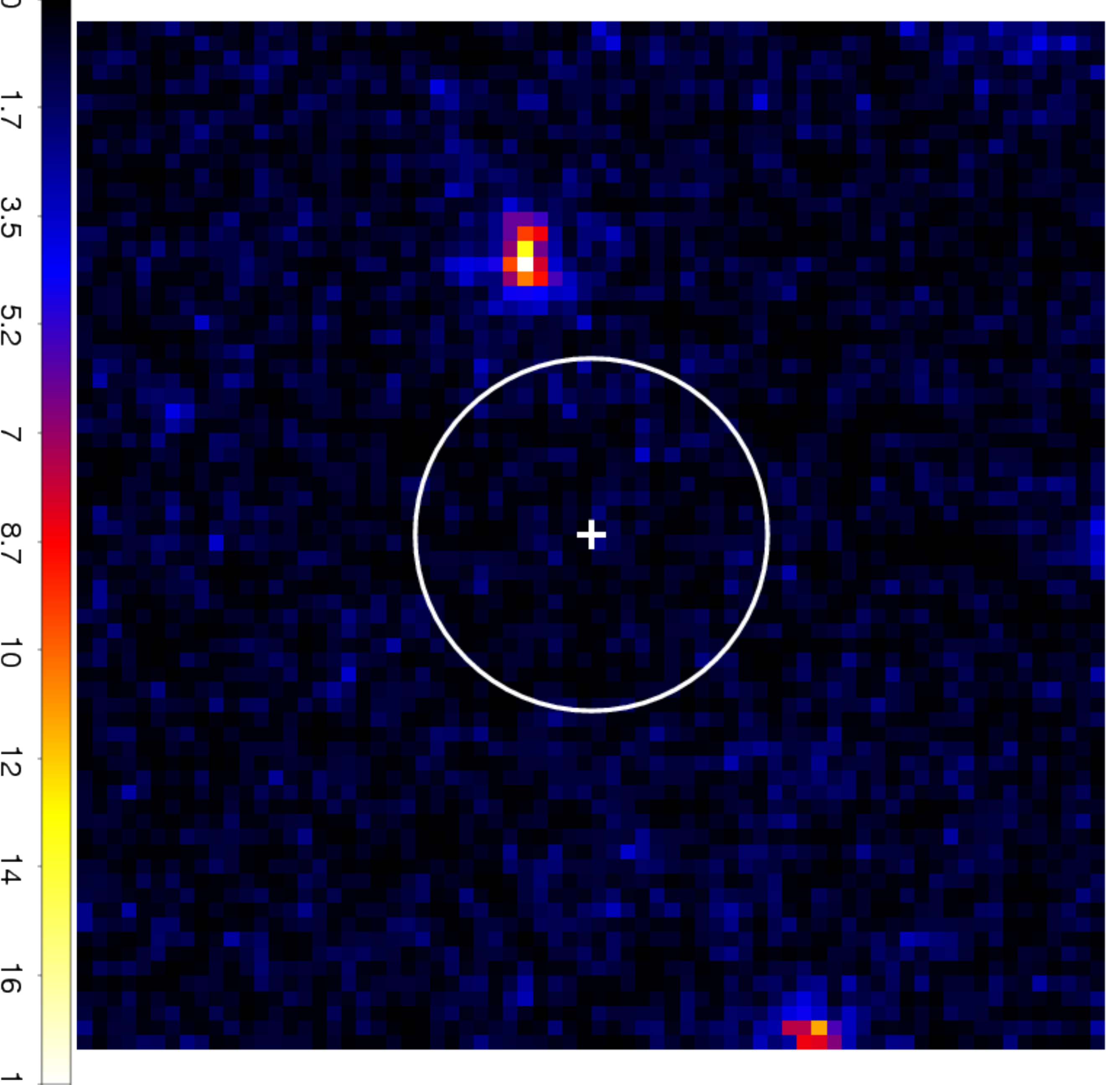}\hfill
\includegraphics[angle=90,width=.33\textwidth]{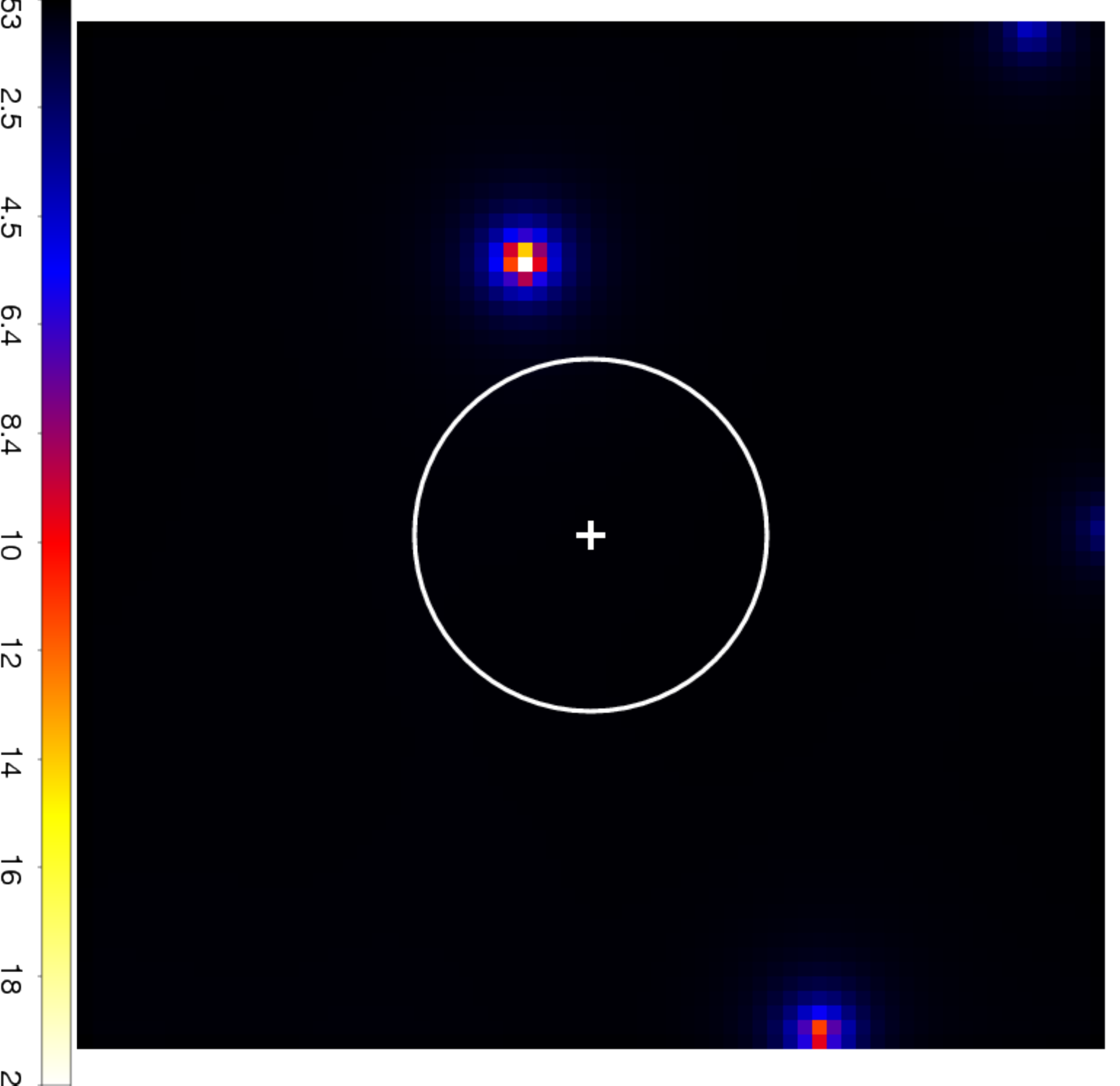}\hfill
\includegraphics[angle=90,width=.33\textwidth]{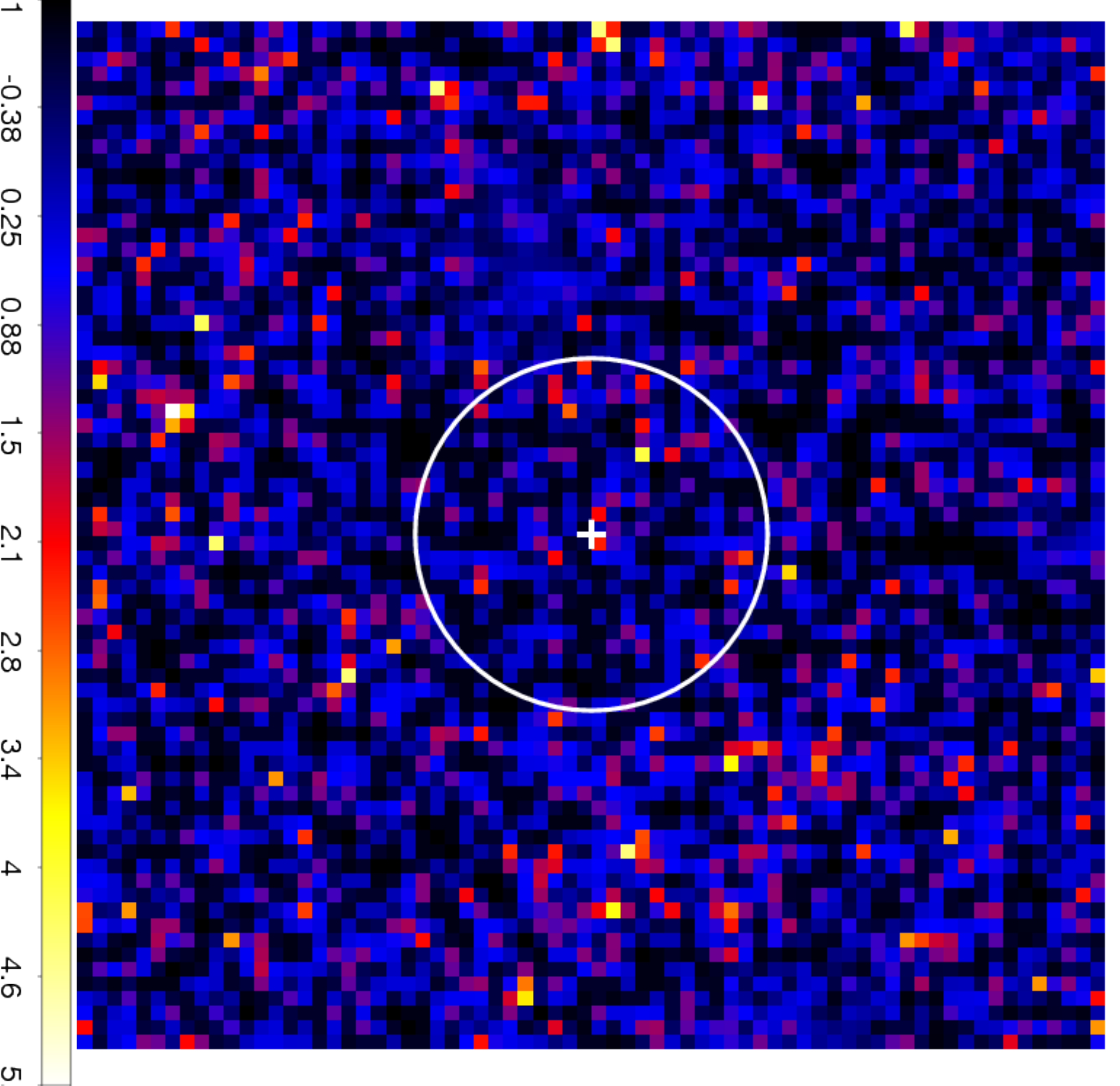}

\caption{7\ensuremath{^{\circ}}$\times$7\ensuremath{^{\circ}} sky, model and residuals maps for Event 37. There is no evidence for new $\gamma$-ray sources beyond the diffuse $+$ 2FGL point sources model used.}
\label{fig:figure3}

 \end{minipage}
\end{figure*}

For the majority of the candidate neutrino events, the residuals maps do not reveal any obvious excess in $\gamma$-ray emission coincident with the neutrino's point of origin. Indeed, the residuals maps for events \#3, \#8, \#13, \#18 and \#37, show no evidence of additional point sources beyond the 2FGL point sources described in the model file. However, the residuals maps for events \#5 and \#23, seen in Figures 2 and 6 respectively, show a clear point source excess. These excesses indicate the presence of a new $\gamma$-ray point source, which is not in the 2FGL catalogue, in the field of view. These new $\gamma$-ray sources will be discussed in detail in \textsection 3.

To search for faint $\gamma$-ray sources associated with the candidate neutrino's origin, that may not be apparent in the residual sky map, a further \textsc{binned} analysis was conducted, with an additional point source included in the model file at the position of the candidate neutrino event. The spectral form of these additional point sources was described by a power law spectrum, $dN/dE = $ A$ \times (E/E_o)^{-\Gamma}$, with both A and $\Gamma$ left free to vary. The ($\alpha_{J2000}$, $\delta_{J2000}$) for this `neutrino' point source were fixed to the values published by the IceCube Collaboration (\citet{icecube1} \& \citet{icecube2}). 

All `neutrino' point sources added to the model file for this additional 70-month \textsc{binned} analysis had a test statistic\footnote{The test statistic, TS, is defined as twice the difference between the log-likelihood of two different models, $TS=2[\text{log} L - \text{log} L_{0}]$, where $L$ and $L_{0}$ are defined as the likelihoods when the source is included or not respectively (\citet{mattox2}).} of 0. As such, no evidence for $1-300$ GeV $\gamma$-ray emission spatially coincident with the ($\alpha_{J2000}$, $\delta_{J2000}$) of the HESE track events was found. Upper limits, at the 95\% confidence level, were therefore calculated for the $\gamma$-ray emission, assuming a point source for the candidate neutrino events. These upper limits are summarised in Table 2. 

\subsection{Icecube neutrino event \#5} \label{sectev5}
In addition to a new $\gamma$-ray source within the field of view studied, the HESE event \#5 has a known $\gamma$-ray bright AGN within the $1\sigma$ uncertainty of the event's position of origin. Located ($1.04\pm0.04$)\ensuremath{^{\circ}} from event \#5 is PKS 0723-008. PKS 0723-008 is a BL Lac object, present in both the 1 year and 2 year \textit{Fermi}-LAT AGN catalogues (1LAC and 2LAC; \citet{1lac}, \citet{acker3}). In our 70-month \textsc{binned} analysis, PKS 0723-008 is detected with a TS of 176.8, a spectral index $\Gamma=2.07\pm0.12$ and a flux of $(9.35 \pm 1.20)\times10^{-10}$ ph cm$^{-2}$ s$^{-1}$. Both PKS 0723-008's 70-month flux and spectral index are consistent with the values published in the 2LAC catalogue. 

To quantify the significance of the spatial coincidence of 2LAC AGN with a single track-like neutrino event, the positions of all 2LAC sources were scrambled in  $\alpha_{J2000}$ and $\delta_{J2000}$, and the new positions were cross-correlated with the $\alpha_{J2000}$ and $\delta_{J2000}$ of all neutrino candidates with $E_{\nu}\geq60$ TeV\footnote{As shown in \textsection4, the background neutrino events considered in this study are expected to concentrate at lower energies, $E_{\nu}<60$ TeV.} considered in this study. Repeated 10,000 times, this approach revealed that the chance probability of having a 2LAC AGN positionally coincident with one of the $E_{\nu}\geq60$ TeV track-like neutrino candidates to be $\sim37$\%. This large probability suggests that the spatial coincidence between event \#5 and PKS 0723-008 is a chance event. 

\begin{figure}
 \centering
\includegraphics[width=60mm]{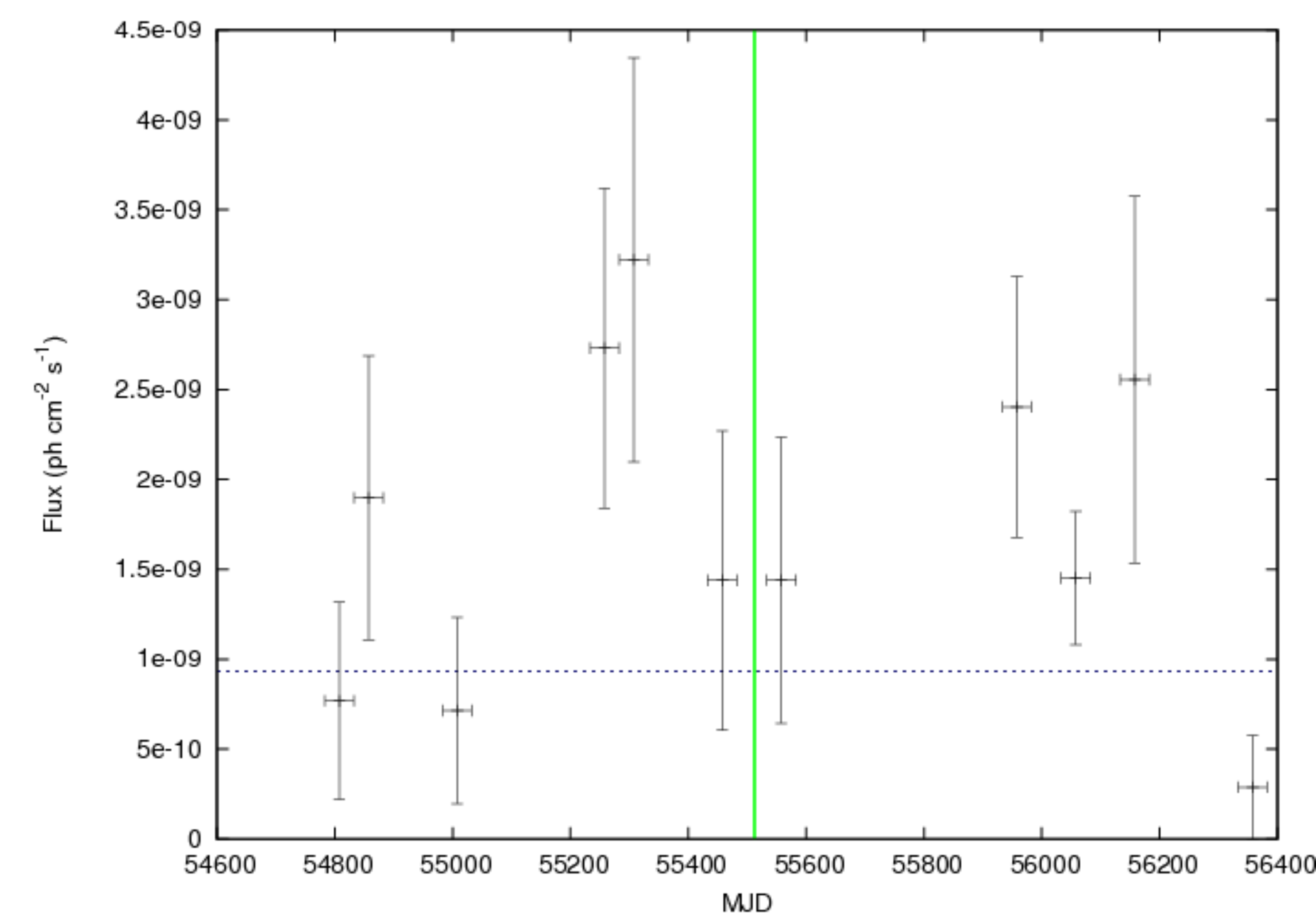}
\caption{The lightcurve of the $1-300$ GeV flux from PKS 0723-008 binned in 50 day periods. The horizontal dashed line shows the flux from the 70-month \textsc{binned} analysis, while the vertical solid line shows the time the \#5 neutrino candidate was detected. While there are periods of enhanced $\gamma$-ray flux from PKS 0723-008 when compared to the 70-month average, there is no obvious flare event associated with the detection of the neutrino event.}
\label{lc}
\end{figure}

Nonetheless, to investigate the temporal characteristics of PKS 0723-008's $1-300$ GeV $\gamma$-ray flux around the time the \#5 neutrino candidate was detected, the data were binned into 50 day periods, with an \textsc{unbinned gtlike} routine applied to each bin separately. Only temporal bins with test statistic $TS>10$ were considered. The resultant lightcurve is shown in Figure \ref{lc}. While the lightcurve does reveal periods of enhanced $\gamma$-ray flux from PKS 0723-008 when compared to the 70-month average, there is no obvious flare event associated with the detection of the candidate neutrino event. Indeed, during the 50-day period in which the neutrino candidate was detected by IceCube, no significant $\gamma$-ray flux was observed.

Phenomenological studies by \textit{Fermi} have found the majority of TeV bright AGN to have $\Gamma < 2$ (\citet{abdo}). Since the 70-month averaged spectrum of PKS 0723-008 is marginally softer than this critical value, we searched for $E_{\gamma}>100$ GeV photons spatially coincident with PKS 0723-008. A closer inspection of the individual photon events reveals the presence of a 115 GeV \textsc{ultraclean} photon within 0\ensuremath{^{\circ}}.1 of PKS 0723-008. Utilising the `diffuse $+$ 2FGL $+$ new source' model, with all model parameters frozen to their 70 month best fit parameters, the \textit{Fermi} tool \textsc{gtsrcprob} calculated the probability that the VHE photon originated from PKS 0723-008 to be 0.97806. Assuming Gaussian errors, this equates to $<2.5\sigma$ significance of the 115 GeV photon originating from PKS 0723-008. There is therefore no evidence of PKS 0723-008 being a source of VHE photons. 

\begin{table*}
 \begin{minipage}{150mm}
   \caption{Summary of upper limits for the $1-300$ GeV $\gamma$-ray flux spatially coincident with the ($\alpha_{J2000}$, $\delta_{J2000}$) of the HESE track events.}
   \begin{center}
     \begin{tabular}{ccccccc} \hline \hline
      IceCube event & Deposited Energy       &  $\alpha_{J2000}$ & $\delta_{J2000}$& Angular error &Northern/Southern & $\gamma$-ray flux upper limit  \\ 
                    &  (TeV)                & (deg)             & (deg)           &  (deg)        &Hemisphere        & (ph cm$^{-2}$ s$^{-1}$)        \\ \hline
      \#3             & $78.7^{+10.8}_{-8.7}$ & 127.9             & -31.2           & $\le 1.4$     &South             & $ 1.89 \times 10^{-10}$        \\ 
      \#5             & $71.4^{+9.0}_{-9.0}$  & 110.6             & -0.4            & $\le 1.2$     &South	     & $ 9.09 \times 10^{-11}$        \\
      \#8             & $32.6^{+10.3}_{-11.1}$& 182.4             & -21.2           & $\le 1.3$     &South             & $ 8.64 \times 10^{-11}$        \\ 
      \#13            & $253^{+26}_{-22}$     & 67.9              & 40.3            & $\le 1.2$     &North             & $ 1.99 \times 10^{-10}$        \\
      \#18            & $31.5^{+4.6}_{-3.3}$  & 345.6             & -24.8           & $\le 1.3$     &South             & $ 1.05 \times 10^{-10}$        \\
      \#23            & $82.2^{+8.6}_{-8.4}$  & 208.7             & -13.2           & $\le 1.9$     &South             & $ 1.32 \times 10^{-10}$        \\
      \#37            & $30.8^{+3.3}_{-3.5}$  & 167.3             & 20.7            & $\le 1.2$     &North             & $ 5.18 \times 10^{-11}$        \\   \hline \hline
    \end{tabular}
  \end{center}
  \label{upperlimits}
\end{minipage}
\end{table*}

\section{New $\gamma$-ray point sources}

The 70-month long exposure analysed in this study revealed two new $\gamma$-ray point sources in the field of view of IceCube's HESE events \#5 and \#23. To investigate the global $\gamma$-ray properties of these new point sources, an additional \textsc{binned} analysis was performed for both of these events, with the model file modified to include an additional point source, described by a power law and spatially coincident with the excesses in residuals map plots of Figures 2 \& 6. 

For the point source in the field of view of the \#5 event, the resultant best fit was found to be:
\begin{equation}
 \dfrac{dN}{dE}= (1.01 \pm 0.24) \times 10^{-14} (\dfrac{E}{6938\text{ MeV}})^{-2.441 \pm 0.189} \nonumber
\end{equation}
\begin{equation}
 \text{ photons cm}^{-2} \text{s}^{-1} \text{MeV}^{-1}
\end{equation}
which equates to an integrated flux of
\begin{equation}
  F_{E_{\gamma}>1\text{ GeV}} = (7.95 \pm 1.23) \times 10^{-10}  \text{ photons cm}^{-2} \text{s}^{-1}
\end{equation}
taking only statistical errors into account. The test statistic value of the best-fit power law was $TS=92.4$, equating to a significance of $\sim9.6\sigma$.

For the point source in the field of view of the \#23 event, the resultant best fit was found to be:
\begin{equation}
 \dfrac{dN}{dE}= (1.761 \pm 0.01) \times 10^{-11} (\dfrac{E}{473\text{ MeV}})^{-2.752 \pm 0.003} \nonumber
\end{equation}
\begin{equation}
 \text{ photons cm}^{-2} \text{s}^{-1} \text{MeV}^{-1}
\end{equation}
which equates to an integrated flux of
\begin{equation}
  F_{E_{\gamma}>1\text{ GeV}} = (1.28 \pm 0.08) \times 10^{-9}  \text{ photons cm}^{-2} \text{s}^{-1}
\end{equation}
again taking only statistical errors into account. The TS value of the best-fit power law was $TS=220.6$, equating to a significance of $\sim14.9\sigma$.

The association of these new $\gamma$-ray sources with known astrophysical sources requires a more precise localisation of the $\gamma$-ray emission. This was done with the \textit{Fermi} tool \textsc{gtfindsrc}, in conjunction with an unbinned exposure map. Using the best-fit model from the 70-month analysis, the \textsc{gtfindsrc} derived ($\alpha_{J2000}$, $\delta_{J2000}$), along with the 1$\sigma$ error radius, for the two new point sources, are given in Table \ref{indexdetails}.

The new $\gamma$-ray point source within event \#23's field of view is spatially coincident with the flat spectrum radio quasar (FSRQ) PKS 1346-112\footnote{We note that the recently published third Fermi source catalogue has found PKS 1346-112 to be $\gamma$-ray bright from 4 years of data (\citet{3fgl}).}. For the new $\gamma$-ray point source within event \#5's field of view, while there is no counterpart within \textsc{gtfindsrc}'s the 1$\sigma$ error radius, the radio source NVSS J$072534+021645$ does lie within the 2$\sigma$ error radius of the new source. This spatial coincidence suggests that this new $\gamma$-ray point source is a blazar. 

\begin{table*}
 \begin{minipage}{150mm}   
  \caption{Summary of ($\alpha_{J2000}$, $\delta_{J2000}$), flux and spectral indices for the new $\gamma$-ray point sources detected in the field of view for event \#5 and \#23. The new point source in the field of view of \#5 is positionally coincident with the radio source NVSS J$072534+021645$ while the point source in the field of view of \#23 is positionally coincident with the AGN PKS 1346-112.}
   \begin{center}
     \begin{tabular}{cccccccc} \hline \hline
      Event \# & $1-300$ GeV Flux                   & Spectral Index  & TS    &$\alpha_{J2000}$ & $\delta_{J2000}$ & 1$\sigma$ error radius & Associated source \\  
               & (ph cm$^{-2}$ s$^{-1}$)            &                 &       &(deg)            & (deg)            & (deg)             &                   \\ \hline
       5       & ($7.95 \pm 1.23$) $\times 10^{-10}$& $2.441\pm0.189$ & 92.4  &111.436          & 2.23478          & 0.039             & NVSS J$072534+021645$ \\ 
       23      & ($1.28 \pm 0.08$) $\times 10^{-9}$ & $2.752\pm0.003$ & 220.6 &207.411          & -11.5502         & 0.031             & PKS 1346-112      \\ \hline \hline
    \end{tabular}
  \end{center}
  \label{indexdetails}
 \end{minipage}
\end{table*}

\section{Discussion}

Of all extraterrestrial neutrino candidates considered in this study, only event \#5 has a $\gamma$-ray point source within the median angular error of the track reconstruction, with the BL Lac object PKS 0723-008 located $\sim1$\ensuremath{^{\circ}} from \#5's origin. However, Monte Carlo simulations revealed the probability that a 2LAC source is positionally coincident with event \#5 to be $\sim37$\%, suggesting that this positional coincidence is merely a chance event. 

In addition to the high probability that a 2LAC source is positionally coincident with a single track-like neutrino event, there are no distinguishing characteristics in the $1-300$ GeV $\gamma$-ray properties, when compared to the other BL Lac objects in the 2LAC catalogue, to suggest PKS 0723-008 is a source of neutrinos. For example, while the 50-day binned lightcurve shown in Figure 8, reveals periods of increased $\gamma$-ray flux $(2-3)$ times greater than the 70-month average, these flare periods do not coincide with the detection of neutrino event \#5. Indeed, during the 50 day period in which the neutrino candidate was observed, no significant $1-300$ GeV $\gamma$-ray emission was observed from PKS 0723-008. Likewise, while PKS 0723-008 does appear to have a fairly hard spectrum, there is no evidence that PKS 0723-008 is a source of VHE photons. So in summary the combination of the insignificant spatial coincidence and the the observed $\gamma$-ray properties suggest that PKS 0723-008 is not the source of the \#5 neutrino candidate. Therefore, this study has found the origin of the HESE track events to be $\gamma$-ray dark in the $1-300$ GeV energy range. 

A possible explanation for the non-detection of $1-300$ GeV $\gamma$-ray emission spatially coincident with the observed neutrino events is that the diffuse neutrino flux observed by IceCube is the aggregate of a very large number of faint neutrino sources which, correspondingly, are faint $\gamma$-ray sources. IceCube’s own point source limits (\citet{icecubePS}) mean that a diffuse astrophysical flux at the level of $10^{-8}$ GeV cm$^{-2}$ s$^{-1}$ sr$^{-1}$, as reported in (\citet{icecube2}), must be distributed over at least O(50) sources. In order to be able to relate the neutrino flux from an individual source to a corresponding photon flux at GeV energies requires detailed modelling which is beyond the scope of this paper. However we note that typical models (\citet{sabaetal}, \citet{doertetal}, \citet{halzenzas}) relating a source's photon and neutrino bolometric luminosities, would suggest that a photon flux, at the level of the upper limits presented in Table 2, is associated with an $E^{-2}$ neutrino flux at least an order of magnitude below the current IceCube point source limits. Thus,  if the TeV photons associated with the neutral mesons are able to quickly cascade down to the GeV energies considered in this dedicated analysis of \textit{Fermi}-LAT observations {\em and} if the typical relationships between the neutrino and photon bolometric luminosities hold, then our photon limits suggest that the diffuse neutrino flux observed by IceCube is drawn from the aggregate of at least O(500) sources.

Alternative explanations could be that the assumption of efficient cascade of TeV photons to GeV energies is not valid, or that the GeV emission is absorbed after leaving the neutrino source or that the neutrino candidates we have studied are in fact residual atmospheric background events. We briefly touch on each of these possibilities below.

(i) The ratio of the neutrino and $\gamma$-ray energies produced through the decay of $\pi^0$, $\pi^+$ and $\pi^-$ mesons is close to unity, such that the $30-250$ TeV neutrinos considered in this study should be accompanied by $\gamma$-rays of a similiar energy. These TeV photons are believed to quickly cascade down to $<$100 GeV energies due to the photon opacity of the emission region (\citet{mannheim}). In the context of the emission region's $\gamma$-ray opacity, the $\gamma$-ray non-detection of the neutrino sources of our study can be interpreted as being due to the opacity being either too small or too large. If the opacity is too low, the TeV $\gamma$-rays are able to freely escape and thus the $\gamma$-ray flux is not able to cascade down to the photon energy range considered in this study. Likewise, if the opacity is too high, the TeV $\gamma$-ray flux would quickly be processed to energies well below the photon energy range considered in this study. Observations with the future Cherenkov Telescope Array\footnote{It is worth highlighting that the upper limits of Table 2 are comparable to the 50 hour (20 GeV - 70 TeV) integrated sensitivity of CTA (\citet{bern}).} or HAWC (\citet{cta}; \citet{hawc}) would shed light onto the former possibility, while detailed multi-wavelength studies would be needed to investigate the latter possibility. 

(ii) In parallel to the effects of internal absorption, the lack of $\gamma$-ray emission associated with the HESE track events could be the result of photon absorption external to the neutrino source. This attentuation primarily occurs through the process of photon-photon pair production ($\gamma\gamma \rightarrow$ e$^+$e$^-$). For $\gamma$-rays, the photon field primarily responsible for this photon-photon attentuation is the Extragalactic Background Light (EBL; e.g \citet{stecker}). A study undertaken by the HESS collaboration found that the intensity of the EBL, at near-infrared/optical wavelengths, is close to the lower limit given by the integrated light of resolved galaxies (\citet{hessebl}). More recently, the \textit{Fermi}-LAT collaboration extended this study using $1-500$ GeV data from the first 46 months of \textit{Fermi}-LAT operation, and drew a similiar conclusion (\citet{fermiebl}). Importantly, the \textit{Fermi}-LAT collaboration also defined a critical photon energy, $E_{crit}$, below which $\leq5$\% of the $\gamma$-ray photons are absorbed by the EBL. The lowest $E_{crit}$ they derived was $\sim30$ GeV for $\gamma$-ray sources in the redshift range $0.5 < z < 1.6$. As such, while invoking $\gamma\gamma$ absorption by the EBL as the primary reason for the $\gamma$-ray non-detection can account for the lack of $E_{\gamma}>30$ GeV flux, it cannot solely explain the lack of $1-30$ GeV flux.

(iii) Finally, we note that IceCube's HESE analysis claimed the discovery of extraterrestrial neutrinos at a $\sim5.6\sigma$ confidence level, with 37 candidate neutrino events detected relative to an expected background of $8.4 \pm 4.2$ cosmic ray muon events and $6.6^{+5.9}_{-1.6}$ atmospheric neutrinos. As such, anywhere between a quarter and two-thirds of the neutrino candidates detected are expected to be background events from cosmic rays interacting with the Earth's atmosphere. These background events are expected to concentrate at lower energies (see Figure 2 of \citet{icecube2}).  In \citet{padovani}, the residual atmospheric background contamination was reduced by only considering the $E_{\nu}\geq 60$ TeV neutrino candidates in their correlation studies. Applying an $E_{\nu}\geq 60$ TeV cut to the track events of our study removes 3 neutrino candidates, events \#8, \#18 and \#37. A closer inspection of the background atmospheric rates\footnote{Here we refer to the atmospheric background as being atmospheric muons and atmospheric neutrinos.} expected in the $30-40$ TeV energy range and declination bands of these removed events, as shown by supplemental Figure 5 of \citet{icecube2}, reveals an expected background of 1-2 events in the 10 \ensuremath{^{\circ}} declination band around events \#8 and \#18 and around 1 event in the 10 \ensuremath{^{\circ}}  declination band around event \# 37, making it rather likely that these neutrino candidates have an atmospheric origin. 

However, inspecting the expected background rates for the $E_{\nu}\geq 60$ TeV neutrino candidates in their respective energy and declination band reveals a smaller number of expected background atmospheric events, ranging from $<0.1$ to $0.25$. In particular for event \#13, which has a deposited energy of $253^{+26}_{-22}$ TeV and a northern hemisphere origin, $\sim0.06$ background events are expected in its declination band during the 988 day IceCube livetime of the 3 year HESE analysis (see supplemental Figure 5 of \citet{icecube2}). This means that, while a contribution from the residual atmospheric background contamination could explain some of the non-detections of possible $\gamma$-ray counterparts to the HESE track events, particularly at lower energies, it cannot explain all of them.

\section{Conclusions}
The angular resolution of a $E_{\gamma}\approx1$ GeV photon detected by the \textit{Fermi}-LAT detector is comparable to the median angular error of IceCube's HESE neutrino candidates detected as track-type events. As such, associating the track-like events of IceCube's HESE analysis with known \textit{Fermi}-LAT detected $E_{\gamma}>1$ GeV sources is simplified greatly when compared to trying to find counterparts to IceCube's shower-like events. Motivated by this, we analysed 70 months of \textit{Fermi}-LAT observations to search for $1-300$ GeV emission spatially coincident with the origin of IceCube's HESE track events. With the exception of event \#5, no $E_{\gamma}>1$ GeV emission spatially coincident with the origin of the neutrino events was found. For these events, 95\% confidence level upper limits were calculated. 

Event \#5 is the only HESE track-like event that has a $\gamma$-ray point source within the median error of its origin, with PKS 0723-008 being separated from the origin by $\sim1$\ensuremath{^{\circ}}. A Monte Carlo study found that the probability of such a spatial coincidence occuring by chance is relatively high, with it occuring in 37\% of the cases we simulated. As such, an insignificant spatial coincidence, combined with the observed $\gamma$-ray properties, suggests that PKS 0723-008 is not the source of the \#5 neutrino candidate.

The study undertaken here has been conducted on the assumption that the intrinsic TeV photons associated with the neutral mesons are able to quickly cascade down to GeV energies without being absorbed entirely. We note that, if this assumption is incorrect, and the TeV photons are able to freely escape from the source, observations with ground-based arrays such as HESS, VERITAS and CTA are required to search for the $\gamma$-ray counterparts to IceCube's neutrino events. However, if the production of these neutrinos is due to transient events, the survey nature of HAWC observations would be required. These observations are strongly encouraged.

\section*{Acknowledgments}

This work was undertaken with the financial support of Durham University and the Marsden Fund Council of New Zealand, and has made use of public \textit{Fermi} data obtained from the High Energy Astrophysics Science Archive Research Center (HEASARC), provided by NASA’s Goddard Space Flight Center. This work has also made use of the NASA/IPAC Extragalactic Database (NED), which is operated by the Jet Propulsion Laboratory, Caltech, under contract with the National Aeronautics and Space Administration. We thank the \textit{Fermi}-LAT collaboration for the quality of the data and analysis tools that were used in this study. Finally we thank the referee for her/his comments and suggestions that have improved the quality and clarity of this paper.

\end{document}